\documentclass[journal]{IEEEtran}
\usepackage[T1]{fontenc}
\usepackage[utf8]{inputenc}
\usepackage{array}
\usepackage{url}
\usepackage{multirow}
\usepackage{graphicx}
\usepackage[numbers]{natbib}
\usepackage[unicode=true]
 {hyperref}

\makeatletter

\providecommand{\tabularnewline}{\\}

\newcommand{\scoping}{\textit{literature}\ \allowbreak\textit{review}}
\newcommand{\exper}{\textit{experiential}\ \allowbreak\textit{review}}

\makeatother

\begin{document}
\title{Review Ecosystems to access Educational XR Experiences: a Scoping
Review}
\author{Shaun Bangay, Adam P.A. Cardilini, Sophie McKenzie, Maria Nicholas
and Manjeet Singh\thanks{Shaun Bangay is with the School of Information Technology, Deakin
University, Geelong, Australia, e-mail: \protect\href{mailto:shaun.bangay@deakin.edu.au}{shaun.bangay@deakin.edu.au}}\thanks{Adam P.A. Cardilini is with the School of Life and Environmental Sciences,
Deakin University, Geelong, Australia, e-mail: \protect\href{mailto:adam.cardilini@deakin.edu.au}{adam.cardilini@deakin.edu.au}}\thanks{Sophie McKenzie is with the School of Information Technology, Deakin
University, Geelong, Australia, e-mail: \protect\href{mailto:sophie.mckenzie@deakin.edu.au}{sophie.mckenzie@deakin.edu.au}}\thanks{Maria Nicholas is with the School of Education, Deakin University,
Geelong, Australia, e-mail: \protect\href{mailto:maria.n@deakin.edu.au}{maria.n@deakin.edu.au}}\thanks{Manjeet Singh is with the School of Information Technology, Deakin
University, Geelong, Australia, e-mail: \protect\href{mailto:maria.n@deakin.edu.au}{manjeet@deakin.edu.au}}}
\maketitle
\begin{abstract}
Educators, developers, and other stakeholders face challenges when
creating, adapting, and utilizing virtual and augmented reality (XR)
experiences for teaching curriculum topics. User created reviews of
these applications provide important information about their relevance
and effectiveness in supporting achievement of educational outcomes.
To make these reviews accessible, relevant, and useful, they must
be readily available and presented in a format that supports decision-making
by educators. This paper identifies best practices for developing
a new review ecosystem by analyzing existing approaches to providing
reviews of interactive experiences. It focuses on the form and format
of these reviews, as well as the mechanisms for sharing information
about experiences and identifying which ones are most effective. The
paper also examines the incentives that drive review creation and
maintenance, ensuring that new experiences receive attention from
reviewers and that relevant information is updated when necessary.
The strategies and opportunities for developing an educational XR
(eduXR) review ecosystem include methods for measuring properties
such as quality metrics, engaging a broad range of stakeholders in
the review process, and structuring the system as a closed loop managed
by feedback and incentive structures to ensure stability and productivity.
Computing educators are well-positioned to lead the development of
these review ecosystems, which can relate XR experiences to the potential
opportunities for teaching and learning that they offer.
\end{abstract}


\section{Introduction}

\subsection{\label{subsec:Rationale}Rationale}

A range of engaging research-derived and commercial interactive games
and virtual reality experiences have been developed to support teaching
and learning. However, none of these are of any significance unless
teachers are supported in selecting applications relevant to their
curriculum topic and integrating them within their teaching plan \citep{ChaconPrado2023}.
Identifying and assessing individual applications is time consuming
so reviews of these applications provide an objective proxy to curate
these experiences. A review ecosystem is the environment that hosts
and maintains these reviews \citep{Bashir2019}.

Applications classed as games already have flourishing review ecosystems
\citep{Eberhard2018,Kasper2019,Lin2018} although these are often
curated to focus on entertainment value and commercial objectives
rather than to identify educationally relevant concepts. Virtual reality
(VR) applications include VR-based games but can include experiences
designed for education and training, or for other purposes such as
tourism or marketing. While some of these might be visible in existing
game review sites, there is currently an opportunity to purposefully
design platforms that host reviews of educational virtual reality
applications to facilitate their identification, adaptation and use
across a range of curriculum topics.
\begin{figure}
\begin{centering}
\includegraphics[width=0.8\columnwidth]{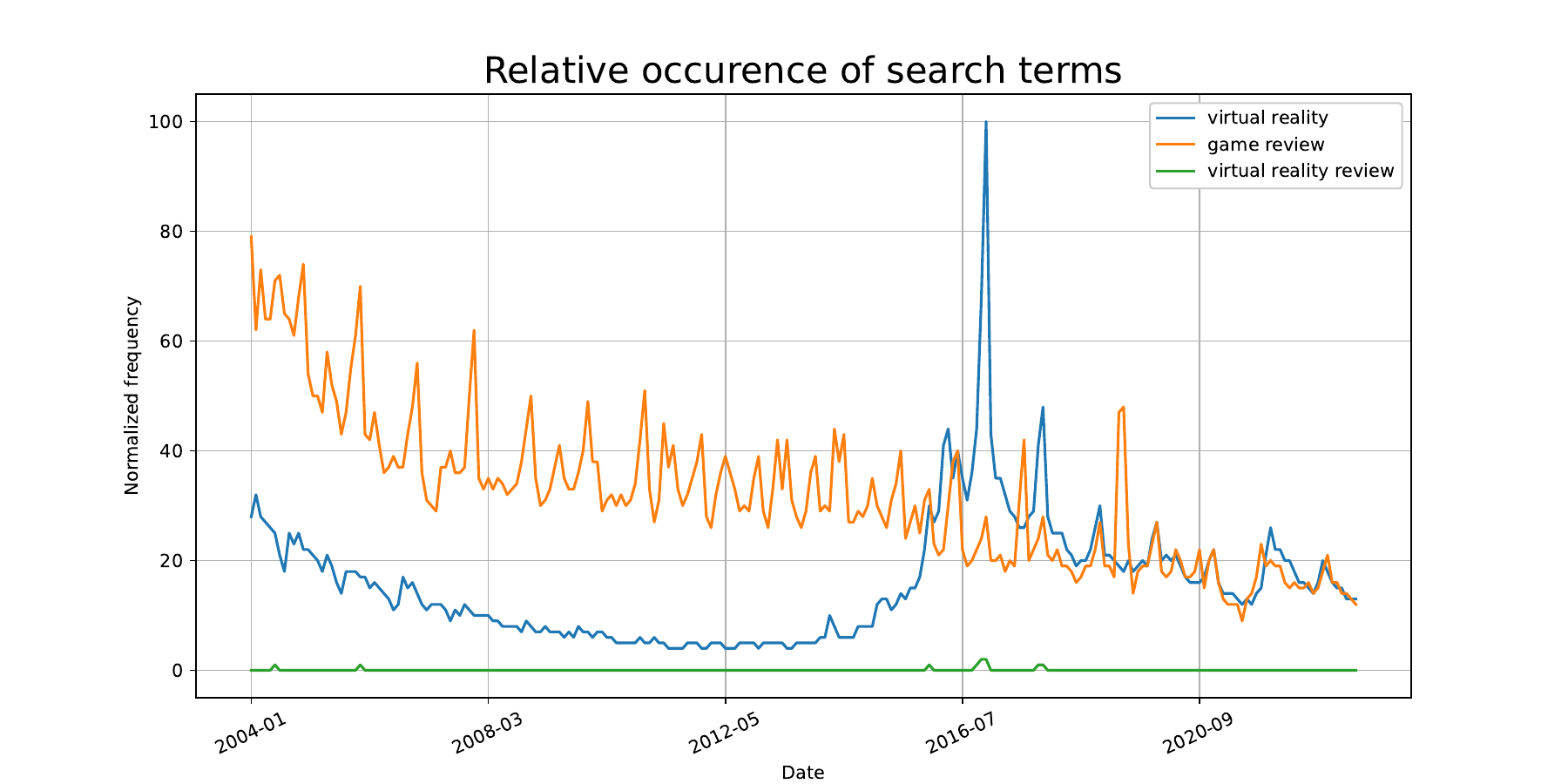}
\par\end{centering}
\caption{\label{fig:Challenges-with-finding}Finding reviews for virtual reality
experiences is challenging as illustrated using Google search trends.
Despite the limited validity of this data source, it does illustrate
that while \textquotedblleft virtual reality\textquotedblright{} is
a well known concept, and \textquotedblleft game review\textquotedblright{}
is also a common search term, the use of \textquotedblleft virtual
reality review\textquotedblright{} is almost nonexistent in comparison.}
\end{figure}
 Figure \ref{fig:Challenges-with-finding} illustrates that despite
interest in virtual reality and an established pattern of use for
game reviews there is a gap and opportunity to create a review ecosystem
that caters for virtual reality reviews.

Academic investigation into serious games \citep{Bedwell2012} does
focus on educational benefits and leads to review environments such
as books \citep{schrier:leg:book} that categorize applications according
to their value in a classroom. Previous systematic literature reviews
\citep{Zheng2021} have identified properties of individual consumer
reviews, including the interactions between the key components of
reviewer, review, recipient, channel and response. We anticipate that
a viable review ecosystem is a complex system that will extend beyond
commercial interests to meet the needs of a diversity of stakeholders
including educators. This review focuses on identifying existing practices
for building and maintaining review ecosystems for interactive applications
so that these can be employed and extended to create specialized review
environments.

A review ecosystem hosts game reviews and facilitates the creation
of new reviews. Review ecosystems for games are well-established,
but those for virtual reality experiences are either a sub-category
of games or are still in their development stages where they may exist
primarily as rating and ranking systems (e.g., for SideQuest\footnote{\url{https://sidequestvr.com/}}).
Educational virtual and augmented reality (\textbf{eduXR}) experiences
have little representation within these systems. The inability of
teachers, and students to assess the appropriateness eduXR experiences
for their learning outcomes represents a significant barrier to the
use of XR in educational settings. An educator who needs to rapidly
identify and deploy such experiences when teaching a specific classroom
topic requires an eduXR review ecosystem with reviews that are created
and formatted for this purpose. \emph{The challenge to be addressed
is to systematically design and build a successful review ecosystem
that exposes the insights captured during the review process, and
that supports stakeholder and community focus beyond gaming and commerce
by reporting on issues relevant to education and XR.} This $\scoping$
catalogs existing best practices to identify review ecosystem components
that can be used as the starting point for achieving this goal.

Teachers face particular challenges in using existing review ecosystems
to identify experiences that support their classes. The benefits of
technology integration, and especially XR \citep{Guilbaud2021,Kuleto2021,Maas2020},
are well documented and include allowing students to intensify their
engagement with concepts \citep{Bernard2018} and engage in both formal
and non-formal learning \citep{Yang2020}. However, after the challenge
of managing equipment, the most significant technology integration
hurdles for teachers are the loss of control, and the need for support
and training in using new experiences \citep{ChaconPrado2023,Hew2006,Francom2019}.
Integration of XR applications, in particular, benefits from social
support mechanisms \citep{Khlaif2024}. Unlike game players who read
reviews at their leisure, most teachers are time poor and need to
find relevant information efficiently \citep{Francom2019}. Teachers
need specific and relevant information to know whether a given experience
can be used within their teaching environment and their available
equipment. The best review ecosystem is one focused on the needs of
teachers and that builds a supportive community of peers \citep{ChaconPrado2023}.
Information in reviews needs to be authentic, accurate and relevant.
The educational merits of particular applications need to have been
validated in practice. The outcomes of this paper are useful for creating
a range of review ecosystems and are applied in section \ref{subsec:Case-study}
to provide the means to design a solution that addresses these requirements.

\subsection{Terminology}

We employ the term eduXR to refer to virtual and augmented reality
experiences (and any blend of these to produce a mixed reality or
extended reality experience \citep{Bangay2022}) that are intended,
or can be repurposed, for educational benefits. Such experiences enhance
the perception of the physical reality experienced by a user with
synthetic virtual content mediated by a computer. The boundary between
XR and other forms of interactive 3D experiences is fluid but for
the purposes of this paper we focus on XR experiences that deliberately
utilize technologies created for the purpose of achieving participant
immersion and that classify themselves as XR.

\begin{figure}
\begin{centering}
\includegraphics[width=0.5\columnwidth]{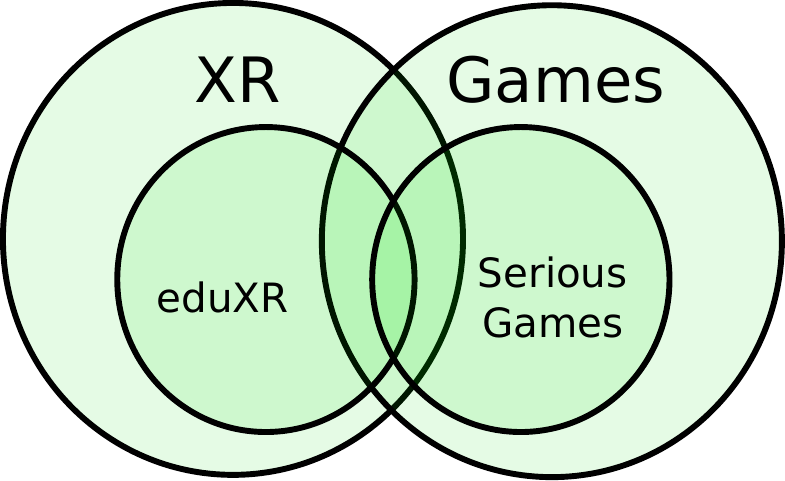}
\par\end{centering}
\caption{\label{fig:The-overlap-between}The overlap between concepts related
to eduXR and serious games.}
\end{figure}

Serious games is a complementary concept to eduXR, with the relationship
being illustrated in Figure \ref{fig:The-overlap-between}.
 The terms are not mutually exclusive as some XR experiences can be
games, although there are other forms (e.g., passive entertainment,
collaboration tools, virtual tourism, educational experiences) that
are not games, just as many games are not intended to be XR experiences.
Serious games provide value beyond entertainment and are frequently
used as a tool to support learning \citep{Bedwell2012}, just as eduXR
focuses on the educational opportunities within XR. Neither the term
seriousXR or eduXR appears to be in widespread use at present although
the latter term is used for some related products.

The word review is used in this paper to describe a report intended
to describe and rate an experience. To avoid confusion this form of
review will be referred to as an $\exper$. This is distinct from
a literature review, such as this scoping review and the systematic
literature reviews in many of the sources that we cite. As such, usage
in the latter context will be described as a $\scoping$. The type
of literature review used for this paper - a scoping review \citep{Tricco2018}
differs from a systematic review in that it aims to identify ``what
is known about a particular concept''. In this case, we aim to identify
and describe those concepts applied to managing game reviews that
can be used as foundational principles for creating a stable and functional
review environment that identifies properties and relationships across
multiple eduXR applications.

\subsection{\label{subsec:Objectives}Objectives}

The goals for this paper are to:
\begin{enumerate}
\item Identify and describe trends related to current practices in preparing,
presenting and maintaining a set of $\exper$s within an $\exper$
ecosystem by using a $\scoping$ methodology. This involves cataloging
existing strategies reported during research into game review ecosystems.
\item Present practices that can be used to establish and enhance the $\exper$
ecosystem for eduXR experiences. Such guidelines would then enable
further research into review ecosystems and how they can take advantage
of XR technologies. These practices and any opportunities identified
can be used to create review ecosystems appropriate to a range of
specialized contexts.
\end{enumerate}
The strategies and practices identified are intended to support the
design and development of a review ecosystem for eduXR that meets
the needs of educators and other stakeholders. The design team need
to consider how best to represent an $\exper$ (the form and format),
ensure the process of creating and presenting a review matches the
needs of the stakeholders (utilization), and must establish the mechanisms
used to produce a stable, complex, control system (ecosystem management).
This $\scoping$ meets these needs by extracting this information
from existing research into review ecosystems.

\section{Method}

This scoping review aims to identify practices used to build and maintain
an $\exper$ ecosystem by identifying practices used in the most closely
related established equivalent; that used for computer games. The
search process identifies reports analyzing existing practices. As
a scoping review \citep{Arksey2005} the focus is to report the extent
of research relevant to creating eduXR review ecosystems, to summarize
research findings and to describe approaches that can be used as a
foundation to further research in this area.

\subsection{Protocol}

The protocol used aligns with the PRISMA process as adapted for scoping
reviews \citep{Tricco2018}. The structure of the scoping review aligns
with the broad objectives of analyzing game review ecosystems and
synthesizing guidelines for developing eduXR review ecosystems \citep{Arksey2005}.

The literature collection (section \ref{subsec:Search}) and analysis
(section \ref{subsec:Synthesis-of-results}) were iterated over two
phases. The first phase refines the search terms and the analysis
process. The second phase performs a clearly defined search and uses
triangulation strategies to confirm adequate coverage. The analysis
extracts information on the strategies used to create review ecosystems
under the categories of \emph{form, usage and ecosystem management}
to categorize the nature of the reviews, the ways in which they are
used, and the ecosystem structures and mechanisms. These topics are
organized to identify trends and presented as options for designing
review ecosystems.

\subsection{Eligibility criteria}

As a scoping review the goal is ensuring that the diversity of the
topic is reflected by capturing insights from all of the sources regardless
of quality \citep{Arksey2005}. All search results were subject to
the same screening based only on the topics covered in the documents.
Each database provides an inherent quality screening based on its
standards for inclusion. A ``review ecosystem'' is not an established
concept so the search process selects research that utilizes collections
of reviews of interactive applications.

\subsection{Information sources}

As the fields of eduXR and computer games are cross disciplinary no
restrictions were imposed on the source of publications used. This
$\scoping$ utilizes two principal multidisciplinary search systems
\citep{Gusenbauer2020}: Web of Science (WoS) and Scopus. The phase
1 search took place in August 2022, with phase 2 updating and extending
this in June 2023.

\subsection{\label{subsec:Search}Search}

The few eduXR review ecosystems that exist are directly derived from
game review sites and so little literature exists that describes the
properties of XR reviews (see section \ref{subsec:Phase-2:-Exhaustive}
below).

\subsubsection{Phase 1: Establish search criteria}

The phase 1 search aimed to trial search terms. The terms used for
each of the databases searched are:

Web of Science: “game OR gaming AND review{*}”, “gaming review”.

Scopus: “game OR gaming AND review{*}”.\footnote{The observant reader will note that the search term should actually
be: “(game OR gaming) AND review{*}”. The consequence of this correction
was tested. In this case the outcome is unchanged because the results
are sorted by relevance which largely compensates for the difference.
The original search actually produces a greater number and variety
of results within the first (n = 350) results, which is helpful for
this exploratory phase.}

The Web of Science search results were grouped in sets of 50 and each
article screened (as per section \ref{subsec:Selection-of-sources}).
The search was halted when a page of 50 entries yielded zero relevant
results (n = 350 entries), yielding 15 relevant entries at the conclusion
of the search. The Scopus search was used to corroborate this, yielding
1 further relevant entry (n = 200).

The word ``review'' identifies any paper that provides a $\scoping$
covering any aspect of gaming. The phase 1 search does include $\scoping$s
that just categorize and analyze games. These papers describe relevant
game properties but are screened out in phase 2 to preserve the focus
on collections of reviews.

\subsubsection{\label{subsec:Phase-2:-Exhaustive}Phase 2: Exhaustive search}

The phase 1 results indicate that relevant papers use the word ``reviews''
(plural) in the title. The following search terms applied to paper
titles provide few useful results:
\begin{itemize}
\item ``virtual~reality'' AND reviews{*}: Web of Science returns 21 results
of which 2 are relevant. The other results relate either to design
reviews (product reviews conducted within virtual reality) or meta-reviews
(reviews of $\scoping$s). Scopus returns 28 documents of which 2
sources are relevant.
\item ``augmented~reality'' AND reviews{*}: Web of Science returns 6
results of which 1 is relevant. Scopus returns 10 results of which
1 is relevant. Discarded results cover the use of AR to present reviews
of other products.
\end{itemize}
Since the number of results was regarded as insufficient to inform
the design of a review ecosystem, the term ``game'' is used as well.
XR applications are typically hosted and reviewed on gaming platforms.
Hence the phase 2 search focused the search using:

Web of Science: ``game{*}'' AND ``reviews'' (in title).

Scopus: TITLE ( \textquotedbl game\textquotedbl{} AND \{reviews\}
).

Web of Science returns 52 results of which 32 pass screening. Scopus
yields 116 documents with 50 relevant, and with 5 papers discarded
as they were identified as earlier versions of later research that
had already been included. This supports the hypothesis that a focus
on game review ecosystems provides greater insight than analysis of
the fledgling XR review systems. 
\begin{figure}
\begin{centering}
\begin{tabular}{|>{\centering}p{0.2\textwidth}|>{\centering}p{0.2\textwidth}|}
\hline 
Phase 1 only & \citep{Koehler2017,Giani2016,YanezGomez2016,Calderon2015,Bas2019,Shiratuddin2011,Petri2017,Coleman2014,Caserman2020}\tabularnewline
\hline 
Phase 1 and 2 & \citep{Lin2018,Wang2020,Kirschner2014,Ho2012,Livingston2011,Kasper2019,Eberhard2018}\tabularnewline
\hline 
Phase 2 only (Web of Science and Scopus) & \citep{Ribeiro2019,Sirbu2016,Viggiato2022,Fong2017,Angelis2021,Urriza2021,Gao2022,Youm2022,Zhu2016,Baowaly2019,Balakrishnan2018,Li2021a,Zhu2015,Faric2019,Straat2017,Vieira2019,Cabellos2022,Tsang2009,Wattanaburanon2016,Kohlburn2022,Guzsvinecz2022,Cox2015,Petrovskaya2022,Kim2014}\tabularnewline
\hline 
Phase 2 only (Web of Science only) & \citep{Ribbens2012}\tabularnewline
\hline 
Phase 2 only (Scopus only) & \citep{Zagal2009,kosmopoulos2020summarizing,Meidl2021,Soetedjo2022,Bond2009,Bian2021,Phillips2021,Yu2021,Cho2020,Busurkina2020,Kwak2020,Strt2017UsingUC,Livingston2010,Petrosino2022,Zubair2021,Santos2019,Zhu2010,Suominen2011,Ahn2017}\tabularnewline
\hline 
Virtual Reality & \citep{Faric2019,Fagernas2021}\tabularnewline
\hline 
Augmented Reality & \citep{Alfaro2021}\tabularnewline
\hline 
Snowballing & \citep{Silva2022,Wang2021,Epp2021,JeongminSeo2023,Deng2023,Boric2023,Tong2021}\tabularnewline
\hline 
\end{tabular}
\par\end{centering}
\caption{\label{tab:Results-returned-from}Results returned from each search
stage.}
\end{figure}

Educational games and XR experiences are not precluded by these search
terms. Topics related to education are identified and extracted during
analysis of each of the papers.

Hence this $\scoping$ analyzes research into gaming review ecosystems
and extrapolates from this to the processes applicable to XR reviews.

\subsubsection{Phase 3: Snowballing and saturation}

\begin{figure}
\begin{centering}
\includegraphics[width=0.4\paperwidth]{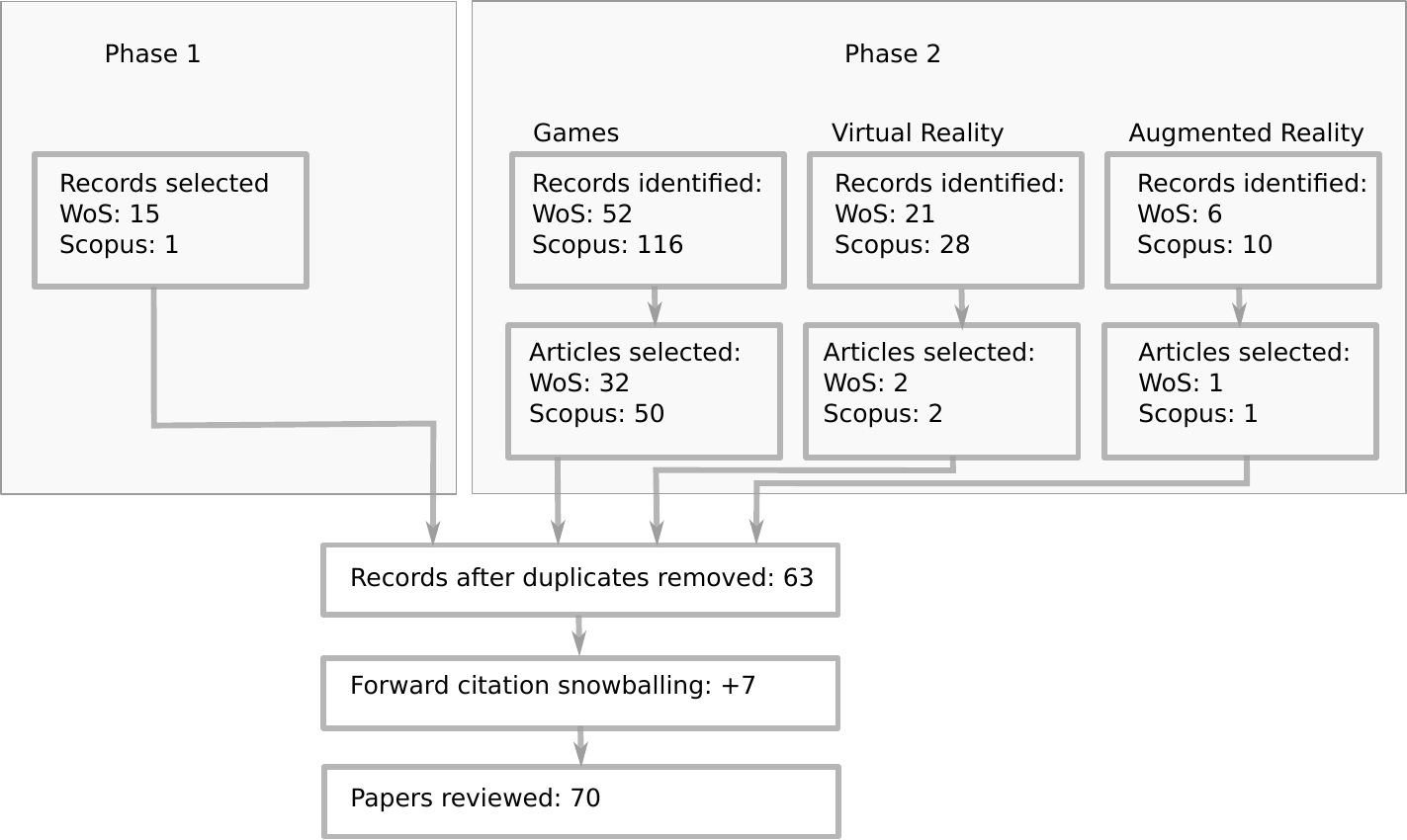}
\par\end{centering}
\caption{\label{fig:Flow-diagram-summarizing}Flow diagram summarizing the
search and screening steps.}
\end{figure}

Saturation is assessed by screening all the papers that cite those
identified, using the links provided in the databases searched. This
snowballing process ensures that recent developments and similar $\scoping$s
are also captured. Snowballing identified a further 7 papers. Reference
lists within the selected papers are examined during analysis and
screened similarly. No additional sources were included although some
references provided insights that are used in the discussion. This
process is summarized in Figure \ref{fig:Flow-diagram-summarizing}
with the papers identified listed in Figure \ref{tab:Results-returned-from}.
 The small number of new papers added at this stage, and the frequent
recurrence of previously found papers suggests that the search process
has identified all relevant sources meeting the screening requirements.
Phase 1 provided a diverse set of sources which contribute to the
breadth of the review. Phase 2 is focused and exhaustive and is dominated
by research into automated mining of review text.

Reviewer feedback suggested a further verification step involving
taking the 1511 results for the WoS title search: \textquotedbl game{*}\textquotedbl{}
AND \textquotedbl review{*}\textquotedbl{} and screening this against
the criteria described using title, keywords and abstract. This produced
the same 32 papers found using the phase 2 search, 4 new papers published
since that search and 1 further paper that used ``game review texts''
rather than ``game reviews'' in the title. This additional triangulation
suggests the literature included is complete. A small number of papers
outside the screening criteria but that elaborate on the identified
themes around educational- and peer-review were also added at this
point.

\subsection{\label{subsec:Selection-of-sources}Selection of sources of evidence}

The phase 1 search sampled the literature by assessing titles to determine
if each article referred to or explored how to conduct an $\exper$
or discussed the features of an $\exper$. The abstract was cross
referenced when additional information was needed. Literature reviews
about gaming and/or associated disorders (e.g., game addiction, sleep
disorders or gambling), or game developers, were excluded from the
final selection. This sample provided guidance on shaping the selection
criteria for the phase 2 search.

In phase 2 the selection criteria applied to screening the paper titles
and abstracts ensure that the paper is focused on games or XR. The
focus of the paper needs to provide insight into how properties of
\emph{collections} of $\exper$s are being analyzed. The paper must
refer to the use of $\exper$s in any form or medium. Generic (i.e.,
unrelated to game properties) discussions (e.g., on social media,
live streams or chat logs), use of game marketing and descriptive
materials, investigation of other forms of engagement measures (e.g.,
psychophysiological measures while playing) and studies that use reviews
without justification are excluded. This excludes analysis of reviews
that describes \emph{only} model fitting (such as text mining, sentiment
distributions e.g., \citep{Secui2016} or natural language properties).
Other forms of product, including other mobile applications, are excluded.
Reviews of other products presented using XR are excluded.

Game evaluations are a form of review that are adequately covered
by other $\scoping$s \citep{Giani2016,YanezGomez2016,Calderon2015}.
Results from phase 1 are retained but are screened out by the phase
2 criteria to focus on collections of reviews. This $\scoping$ is
lenient in screening and aims to include articles if they can contribute
new insights into managing a review ecosystem. Since no meta-analysis
is required rigorous sampling is not necessary.

\subsection{Data items}

The focus of the analysis is to identify the current state of research
into review ecosystems. The information required, as defined in section
\ref{subsec:Objectives}), specifies the following topics as necessary
to guide the design and development of review ecosystems:
\begin{enumerate}
\item \textbf{Form~of~the~$\exper$}: This category covers the way information
in the review is presented, and the trade-off between presenting a
review that is interesting, informative and specific to the eduXR
experience against populating a table of standardized entries that
can easily be searched or compared.
\begin{itemize}
\item \textbf{Format}: The way in which the review is presented (e.g., written
documents, video reviews).
\item \textbf{Template}: Do reviews follow a consistent structure and, if
so, what are the elements of this structure.
\item \textbf{Focus}: The motivation for doing the review.
\item \textbf{Field}: EduXR overlaps with several other fields, including
simulation, games, and serious games. The material reviewed may concentrate
on one or more of these.
\end{itemize}
\item \textbf{Review~utilization}: This category identifies the ways in
which the review is useful. This is inferred from hints provided in
each source where it is not explicit.
\begin{itemize}
\item \textbf{Target~audience}: Who is the review intended for?
\item \textbf{Usage}: What the review is used for, and how it is used for
that purpose.
\item \textbf{Education}: How reviews contribute educational value and the
educational aspects that are measured in a review.
\item \textbf{Mechanics}: How the review is actually created.
\end{itemize}
\item \textbf{Ecosystem~management}: Individual reviews are expected to
have significantly more value when they exist in a shared environment
(e.g., by using them to compare different eduXR experiences).
\begin{itemize}
\item \textbf{Structure}: The components of the ecosystem (e.g., groupings
of stakeholders).
\item \textbf{Incentives}: The mechanisms employed to ensure new eduXR experiences
are reviewed, and that the reviews follow conventions and standards.
\item \textbf{Value}: The rewards and the costs associated with the review
ecosystem. The hypothesis being assessed is that any stable ecosystem
should have rewards exceed costs, and that all stakeholder groups
should perceive a net gain.
\item \textbf{Quality}: Mechanisms to compare or evaluate reviews.
\item \textbf{Environment}: The environment in which the review lives, including
the sites that host them.
\item \textbf{Viability}: The lifespan and ongoing viability of a review
ecosystem.
\item \textbf{Challenges}: The difficulties associated with the creation
of reviews and management of review ecosystems.
\end{itemize}
\end{enumerate}
The selected papers were reviewed to identify any information relevant
to each of the topics. A spreadsheet was prepared with a column for
each topic. The material from each paper that is relevant to each
topic is recorded in the spreadsheet. The analysis is performed by
one member of review team reviewing all papers to ensure consistency
in extracting information from the papers. The information extracted
is then reviewed by all the other authors and the different options
are listed in section \ref{sec:Analysis}.

\subsection{\label{subsec:Synthesis-of-results}Synthesis of results}

Synthesis identifies the trends or alternative approaches for each
topic presented in section \ref{sec:Results}. These present the options
available when designing and implementing a new review ecosystem intended
to achieve a particular purpose. Subsequent discussion points are
approved by all authors and used to identify opportunities to focus
further research to advance best practice in developing $\exper$s
for eduXR experiences. 

\section{\label{sec:Analysis}Analysis}

\subsection{Review ecosystems}

What is a review? The concept of an $\exper$ differs widely across
the literature that was surveyed. The search terms used on academic
literature databases identify many systematic literature reviews during
phase 1. The acceptance criteria require that the papers be relevant
to gaming reviews with the consequence that several papers apply the
term to both forms of review within a single paper. A review can be
conducted both as an $\exper$: a document describing particular properties
of one game \citep{Calderon2015,Eberhard2018,Ho2012,Kasper2019,Koehler2017,Lin2018,Livingston2011,Wang2020},
but also as a $\scoping$ which identify common trends across several
sources \citep{Petri2017,YanezGomez2016}. The latter interpretation
identifies a potential gap in the game review ecosystem as typical
game reviews describe only a single game. Other forms of review are
possible, e.g., heuristic evaluation \citep{Livingston2010}.

Data mining \citep{Wang2020,Eberhard2018,Viggiato2022,Fong2017,Urriza2021,Youm2022,Zhu2015,Vieira2019,Wattanaburanon2016,Kwak2020,Deng2023}
is used to identify features of game reviews that predict game properties
such as ratings. An opportunity still exists to compare trends and
commonalities across different games based on such features. Game
reviews use specific terminology, humour and sarcasm, and reviewers
have varied motivations and incentives, which provides a challenge
when distilling information from collections of reviews \citep{Eberhard2018,Viggiato2022,Angelis2021,Santos2019}.
This is balanced by the reward of greater value resulting from the
collective value of the review ecosystem \citep{Urriza2021,Kohlburn2022}.
Review ecosystems equalize the power imbalance between large commercial
operations and individual players \citep{Youm2022}, and provide an
opportunity to reason about and compare reviews \citep{Baowaly2019,Ribbens2012,Guzsvinecz2022,Fagernas2021,Petrosino2022,Bian2021,Kwak2020,kosmopoulos2020summarizing,Santos2019,Suominen2011,Zagal2009,Bashir2019,Wang2021,Zheng2021,SouzaGoncalves2020}.

The purpose of a review varies. Reviewers can have specific agendas
such as: providing a description and recommendation for a game \citep{Ho2012,Wang2020,Cho2020,Meidl2021,JeongminSeo2023},
presenting personal experiences \citep{Ribeiro2019,Balakrishnan2018},
causing game sales \citep{Zhu2010}, identifying a relevant educational
resource \citep{Caserman2020}, reporting on educational properties
of a game \citep{Giani2016} or how the game functions for learning
particular topics (e.g., software project management \citep{Calderon2015},
computing education \citep{Petri2017} or employee selection \citep{alQallawi2021}),
the game's value as an assessment platform \citep{Bas2019,Calderon2015,alQallawi2021,Barr2017,Friedrich2019},
measuring properties and quality of the game \citep{Eberhard2018,Kasper2019,Koehler2017,Lin2018,Caserman2020,Sirbu2016,Fong2017,Zhu2016,Cabellos2022,Phillips2021,Ahn2017,Bond2009,Huang2015,Deng2023}
or as a quality control process \citep{Shiratuddin2011}, providing
feedback to game designers and developers \citep{Coleman2014,Faric2019,Straat2017,Yu2021,Busurkina2020,Silva2022,Yu2023},
informing purchase decisions \citep{Livingston2011,Urriza2021,Cox2015,Kim2014},
understanding the player and their experience \citep{Kirschner2014,Gao2022,Tsang2009,Alfaro2021,Soetedjo2022,Zubair2021,Strt2017UsingUC,Epp2021,Boric2023,Tong2021,Guzsvinecz2023}
or assessing and influencing the play experience \citep{YanezGomez2016,Li2021a,Wattanaburanon2016,Petrovskaya2022,Philp2024}.
A review is a way of experiencing an intangible product that cannot
be tested before purchase \citep{Eberhard2018} and bookends the game
experience with anticipation and priming prior and sharing afterwards.

\subsection{\label{subsec:Formofthereview-1}Form~of~the~review}

W e investigate the properties of individual reviews to develop insight
into the mechanisms used to create and manage review ecosystems. Specifically
we identify how reviews are presented, what information they contain
and what they provide.

\subsubsection{Format}

The form and format of reviews reported, as summarized in Figure \ref{tab:Format-1},
is typically as a written document although these exist in several
forms. Several sources review and analyze academic literature in the
form of case studies, and these individual case studies are reviews
of an experience. The structure of a case study is, however, dominated
by formats dictated by the case study process rather than the need
to consistently report on the system being studied. While most written
reviews contain at least some free-form text a few sources use, or
recommend using, particular fields as a template for providing reviews.
Game review sites usually provide limited facilities for formatting
the free-form review text \citep{Kasper2019} rather than supporting
a recommended document structure with consistent headings. Review
writing varies in rigour from formal reports produced by professional
reviewers \citep{Cox2015,Zagal2009} to written reviews created by
other members of the community, and then to short informal comments
provided on social media platforms \citep{Silva2022}. Allowing only
constructive feedback enhances the value of the review \citep{Barr2017}.

Most of the reviews selected for analysis reflect a single culture
and language. This choice conveniently simplifies any processing of
the natural language text. A few studies do explicitly analyze reviews
presented in a range of languages \citep{Viggiato2022} and investigate
whether cultural conventions influence the way information is expressed
in reviews \citep{Tsang2009} and consequently whether it is meaningful
to compare scores generated for these reviews.

Reviews in other formats are comparatively rare with surprisingly
few mentions of video reviews \citep{Lin2018,Urriza2021,Zheng2021}
given the popularity of social media sites for video sharing and streaming
of game play. Content created in 3D (in this case architectural models
\citep{Shiratuddin2011} but relevant to games and XR) could be reviewed
in the same 3D format using collaborative annotation in a shared virtual
space. A further non-traditional form of review \citep{Kirschner2014}
involves deriving insight into game play by mediating the game experience
through recordings of play and discussion with a researcher who then
reports on the outcome. This latter process is distinctive in that
the review is not constructed directly by the player, and that the
review is linked to a particular play session.

\begin{figure}
\begin{centering}
\begin{tabular}{|>{\raggedright}p{0.4\textwidth}|}
\hline 
{\footnotesize{}Review formats}\tabularnewline
\hline 
\hline 
{\footnotesize{}Written document (academic paper): \citep{Bas2019,Calderon2015,YanezGomez2016}}\tabularnewline
\hline 
{\footnotesize{}Written document (fixed format/template): \citep{Kasper2019,Petri2017,Wang2020,Coleman2014,Caserman2020,Wattanaburanon2016,Friedrich2019}}\tabularnewline
\hline 
{\footnotesize{}Written document (unrestricted prose): \citep{Eberhard2018,Ho2012,Kasper2019,Koehler2017,Lin2018,Livingston2011,Giani2016,Sirbu2016,Viggiato2022,Fong2017,Gao2022,Youm2022,Baowaly2019,Ribbens2012,Balakrishnan2018,Li2021a,Zhu2015,Faric2019,Straat2017,Vieira2019,Cabellos2022,Kohlburn2022,Guzsvinecz2022,Petrovskaya2022,Kim2014,Fagernas2021,Alfaro2021,Soetedjo2022,Petrosino2022,Phillips2021,Bian2021,Yu2021,Zubair2021,Kwak2020,Cho2020,kosmopoulos2020summarizing,Busurkina2020,Santos2019,Ahn2017,Strt2017UsingUC,Meidl2021,Zagal2009,Silva2022,Huang2015,Wang2021,Epp2021,JeongminSeo2023,Deng2023,Boric2023,Tong2021,SouzaGoncalves2020,alQallawi2021,Chand2022,Yu2023,Guzsvinecz2023,WangReview2008,Barr2017}}\tabularnewline
\hline 
{\footnotesize{}Other: video \citep{Lin2018,Urriza2021,Zheng2021},
process using video and screen recordings and interviews \citep{Kirschner2014},
images \citep{Suominen2011,Zheng2021}, 3D virtual environment \citep{Shiratuddin2011},
numeric score/rating \citep{Ribeiro2019,Cox2015,Livingston2010,Bond2009,Zhu2010,Balietti2016,BRYAND.2006}}\tabularnewline
\hline 
\end{tabular}
\par\end{centering}
\begin{centering}
\par\end{centering}
\caption{\label{tab:Format-1}Format: the forms and formats of $\exper$s}
\end{figure}

\subsubsection{\label{subsec:Template-1}Template}

Every source identifies particular elements of a review that should
be common to all reviews. These relate to the description of the product,
and include the name, genre, and an aggregation of the previous review
ratings. The remaining categories of review element vary depending
on the context and are summarized in Figure \ref{tab:Template-1}.

Player properties are provided by dedicated game review sites and
report the qualifications of the player (who is assumed to be the
author of the review). Experience is measured in terms of previous
reviews created and by hours of game play, reflecting the extent to
which a review represents first impressions or a level of competence
in the game.

A significant category is the elements that describe properties of
the game. This is a rich and diverse topic and the merits of these
properties are better suited to literature reviews that specialize
in game analysis. Professional reviewers provide detailed and objective
reports while amateurs present personal and emotional reviews \citep{Santos2019}.
Several sources make explicit reference to the quality of the game.
Since a review can be expected to serve as a quality indicator, this
particular game property is discussed in its own category.

While not all reviews are expected to focus on educational applications,
there are sufficiently many represented to justify categories relating
to these review topics. These topics describe the educational content,
and the degree to which the experience supports learning outcomes.
Gamification terms are frequent where the experience is intended to
provide motivation, engagement, clear goals, and feedback to support
learning. The use of the game as an assessment tool merits its own
category which concentrates on measuring achievement before and after
the experience.

Reviews also serve to provide information about the developer of the
game although the sole source \citep{Coleman2014} assumes this information
is provided by the developers themselves.

The review fields dictate the way the review is conducted. For example,
reviews may be conducted at key stages of exposure to the experience
\citep{Bas2019}, or be required to provide information to validate
the rigour of the review \citep{Petri2017}. Review analysis shows
that the structure of the review correlates with quality of the experience.
For example, the format differs between positive and negative reviews
\citep{Lin2018}.

Several sources reason explicitly about the structure of the review
and what information it should contain. This is particularly relevant
when considering reviews published as case studies, and the need to
meet acceptance criteria before these can be utilized within a systematic
review \citep{Calderon2015,Koehler2017,Giani2016,YanezGomez2016}.
The analysis \citep{Alfaro2021} classifies the focus of a review
into the categories of: information giving, information seeking, problem
solving and requesting feature changes. Correlations can exist between
the topics presented. For example, negative reviews often focus on
aspects of design rather than technical problems \citep{Petrosino2022}.
Amateur reviews tend to be produced in isolation, and miss the opportunity
to present similarities to other games, or to employ a standardized
vocabulary to describe game concepts \citep{Cho2020}. In contrast,
professional reviewers produce many reviews and are able to draw connections
to previous versions and other games \citep{Suominen2011}.

The lack of a consistent review template \citep{Bashir2019} has the
consequence that analysis of sets of reviews either concentrates on
just the numerical scores (ignoring the text), performs manual qualitative
analysis on a small number of reviews, or employs automated natural
language processing to identify themes as an initial step \citep{Silva2022}.

\begin{figure*}
\begin{centering}
\begin{tabular}{|>{\centering}p{0.15\textwidth}|>{\raggedright}p{0.6\textwidth}|}
\hline 
{\footnotesize{}Information category} & {\footnotesize{}Topics}\tabularnewline
\hline 
\hline 
{\footnotesize{}Product metadata} & {\footnotesize{}application domain, types of game \citep{Calderon2015,Chand2022},
recommendation \citep{Gao2022,Eberhard2018,alQallawi2021}, author
ID, got game for free, got game before release \citep{Eberhard2018},
name of game \citep{Guzsvinecz2022,Ho2012}, category, version, ranking
(in downloads), star ratings, number of reviews \citep{Ho2012}, price
\citep{Ho2012,Zheng2021}, game information \citep{Koehler2017},
genre \citep{Ahn2017,Suominen2011}, system requirements \citep{Ahn2017,Suominen2011},
brief text overview, awards received, link to site (URL), target age
group \citep{Coleman2014}, tags \citep{Ahn2017}, community size,
special events, time to win \citep{Philp2024}}\tabularnewline
\hline 
{\footnotesize{}Player/reviewer properties} & {\footnotesize{}time played, products owned \citep{Eberhard2018,Cabellos2022,Guzsvinecz2022,Guzsvinecz2023,Philp2024},
playing hours \citep{Lin2018,Deng2023}, recommended/not recommended,
number of played games, number of reviews posted, number of hours
playing before writing a review, negativity of reviews \citep{Lin2018},
details of the subjects using the experience, details of reviewer
\citep{YanezGomez2016}}\tabularnewline
\hline 
{\footnotesize{}Review content} & {\footnotesize{}description \citep{Zagal2009}, pros, cons, suggestions,
bug reports \citep{Lin2018,Urriza2021,Wattanaburanon2016}, rating
(score) \citep{Kasper2019,Livingston2011,Faric2019,Tsang2009,Petrovskaya2022,Kim2014,Fagernas2021,Yu2021,Santos2019,Strt2017UsingUC,Meidl2021,Huang2015,Guzsvinecz2023},
information for other users \citep{Zubair2021}, technical, design
and service issues \citep{Alfaro2021,Zagal2009}, how to maximize
value \citep{Zagal2009}}\tabularnewline
\hline 
{\footnotesize{}Gaming properties} & {\footnotesize{}acceptance \citep{Calderon2015}, accessories \citep{Wang2020},
achievement \citep{Wang2020}, action language \citep{Koehler2017},
assessment \citep{Koehler2017}, audio \citep{Kohlburn2022,kosmopoulos2020summarizing,Suominen2011,Zhu2010,Boric2023},
background \citep{Boric2023}, balance \citep{Coleman2014}, conflict/challenge\citep{Koehler2017,Giani2016,Petri2017,Tong2021},
content \citep{Epp2021}, characters \citep{Kohlburn2022,Deng2023,Boric2023},
control\citep{Koehler2017,Petri2017,Epp2021}, crashes \citep{Epp2021,Deng2023},
customization \citep{Zagal2009}, educational balance \citep{Caserman2020},
effect on user motivation\citep{Calderon2015}, engagement \citep{Petri2017},
enjoyment\citep{Calderon2015,Petri2017}, environment\citep{Koehler2017,Zagal2009},
experience \citep{Busurkina2020}, feedback \citep{Coleman2014,Boric2023},
fit for purpose \citep{Coleman2014}, flow \citep{Giani2016}, fun
\citep{Giani2016,Petri2017}, game design\citep{Calderon2015,Coleman2014,Zagal2009,Boric2023},
game fiction\citep{Koehler2017,Wang2020,Coleman2014,Kohlburn2022,Zagal2009,Boric2023,SouzaGoncalves2020},
game play \citep{kosmopoulos2020summarizing,Suominen2011,Zhu2010,Zagal2009,Epp2021,SouzaGoncalves2020},
immersion\citep{Koehler2017,Giani2016,Petri2017}, influence \citep{Busurkina2020},
mechanics \citep{Kohlburn2022}, motivation \citep{Coleman2014},
multiplayer \citep{Tong2021}, performance\citep{Calderon2015,Epp2021},
physics \citep{Tong2021}, playability\citep{Calderon2015,Kohlburn2022},
player experience \citep{Caserman2020}, price \citep{Kohlburn2022,Zagal2009,Boric2023},
progress \citep{Busurkina2020}, realism \citep{Petri2017}, recommendation
\citep{Petri2017}, rules/goals, \citep{Koehler2017,Tong2021}, social
interaction \citep{Petri2017,Wang2020,Kohlburn2022,Busurkina2020,Zagal2009},
uniqueness \citep{Zhu2015,Kohlburn2022}, usability \citep{Calderon2015,Koehler2017,Giani2016,Petri2017,Zagal2009},
variety \citep{Zagal2009,Tong2021}, visual/value \citep{Wang2020,kosmopoulos2020summarizing,Suominen2011,Zhu2010,Busurkina2020,Boric2023,SouzaGoncalves2020}}\tabularnewline
\hline 
{\footnotesize{}Game quality} & {\footnotesize{}quality characteristics/measures \citep{Calderon2015,Ribbens2012},
value judgement, comparisons \citep{Koehler2017}, quality \citep{Petri2017,Caserman2020}}\tabularnewline
\hline 
{\footnotesize{}Educational content and value} & {\footnotesize{}educational elements, learning outcomes, engages users,
achieving intended result, social impact, affect on cognitive behaviour
\citep{Calderon2015}, learning, social interaction, relevance, goal
clarity, motivation \citep{Giani2016} student learning, quality of
instruction, motivation, instruction \citep{Petri2017}, use for training,
context, elements of the game \citep{Coleman2014}, serious/goal relevance
(characterizing goal, clear goal, indispensability of goal, content
correctness, feedback on progress, reward, proof of effectiveness)
\citep{Caserman2020}, rubric \citep{Friedrich2019}}\tabularnewline
\hline 
{\footnotesize{}Educational assessment} & {\footnotesize{}categories of assessment (summative, formative, end-of-game,
stealth, scoring, external) \citep{Bas2019}, learning (competence
before and after playing) \citep{Petri2017}}\tabularnewline
\hline 
{\footnotesize{}Review text features} & {\footnotesize{}number of paragraphs, readability, sentiment, time
span, similarity \citep{Eberhard2018}, fun, information richness,
perceived value, after sales support, stability, challenge, expectation,
promotion, online community, accuracy, special event, style of game,
innovation, sustainability \citep{Ho2012}, writing style features,
content features \citep{Kasper2019}, emotion \citep{Cabellos2022},
language \citep{Guzsvinecz2022,Yu2021}}\tabularnewline
\hline 
{\footnotesize{}Developer feedback} & {\footnotesize{}development methodology, challenges, authenticity,
learning considerations, team composition (all stakeholders), deployment,
hardware impact, alternative uses for the game \citep{Coleman2014,Yu2021},
discussion about design decisions \citep{Zagal2009}, developer response
to review \citep{alQallawi2021}, coding standard \citep{WangReview2008}}\tabularnewline
\hline 
{\footnotesize{}Review processes} & {\footnotesize{}structured as: phases of play (preparation, introduction,
interaction, conclusion) \citep{Bas2019}, based on review outcome
(positive/negative), indie/non-indie, early access, free to play,
when posted \citep{Lin2018}, factors evaluated, research design,
methods, data collection, sample sizes, replication, data analysis
\citep{Petri2017}, social influence (within the review process) \citep{Wang2020}}\tabularnewline
\hline 
{\footnotesize{}Meta review (properties of the review)} & {\footnotesize{}how evaluation is conducted, population size \citep{Calderon2015},
value/helpfulness of the review \citep{Baowaly2019,Guzsvinecz2022,kosmopoulos2020summarizing,Santos2019},
humour rating \citep{Yu2021}, game information, text of the review
\citep{Koehler2017}, clearly defined approach, with empirical evaluation
\citep{Giani2016}, review format (new structure, using templates,
adapting templates, custom) \citep{YanezGomez2016}, biased or fake
\citep{Bian2021}, reviewer history \citep{Wang2021}}\tabularnewline
\hline 
{\footnotesize{}Other} & {\footnotesize{}compliance with various codes, maintenance requirements
\citep{Shiratuddin2011}, therapeutic value \citep{Phillips2021}}\tabularnewline
\hline 
\end{tabular}
\par\end{centering}

\caption{\label{tab:Template-1}Template: elements expected in an $\exper$}
\end{figure*}

\subsubsection{\label{subsec:Focus-1}Focus}

The reasons for performing the review vary as much as the structure
and format of reviews. The obvious purpose of a game review is to
identify pleasurable experiences \citep{Zagal2009}, rate game features
\citep{Guzsvinecz2022} and inform other players of the nature of
the game \citep{Zhu2016} but even these goals has their subtleties.
The reviews may be directed at peers \citep{Koehler2017,Balietti2016,Barr2017},
in which case communication is targeted at others with equivalent
interests and represents a voluntary contribution to a community of
like-minded individuals \citep{Gao2022,Zhu2015,Silva2022,Tong2021,Bianchi2018,Garcia2017,Garcia2020}.
Reviews are a way to achieve recognition and standing within such
communities \citep{Baowaly2019,Petrovskaya2022,Deng2023}, and provide
a forum to relate game play themes to the reviewer's own lived experience
\citep{Kohlburn2022,Phillips2021,Guzsvinecz2023} or personal agenda
\citep{Fagernas2021,Busurkina2020,Zhu2010}. Reviews might be created
just because of a need for the insights that they provide \citep{Bashir2019}.
Experienced gamers \citep{Wang2021} and professional reviewers produce
longer and more complex reviews \citep{Viggiato2022} and describe
the game in the context of its predecessors \citep{Suominen2011}.
Reviews can influence the way in which a game is perceived, colour
the play experience \citep{Livingston2011} and produce more sophisticated
and aware players \citep{Straat2017}. Casual reviewers often provide
reviews for old games, suggesting an element of nostalgia \citep{Santos2019}.

Reviews influence purchasing decisions so review ecosystems develop
that facilitate this \citep{Ho2012,Kasper2019,Wattanaburanon2016}.
From this perspective, reviews are tools to influence purchase decisions
\citep{Urriza2021,Tsang2009,Kim2014,kosmopoulos2020summarizing,Zheng2021}.
Reviews provide a mechanism to gain insights into intangible products,
such as games, where it is not possible to test the product before
buying \citep{Youm2022}. Reviews are more informative than the game
descriptions provided by publishers \citep{Cho2020}, and provide
balance to the marketing efforts of large companies \citep{Wang2021}
by allowing individuals to reach large numbers of players \citep{Tong2021}.
Reviewers act as gatekeepers to ensure quality \citep{BRYAND.2006}.
Positive reviews correlate to the success of a game with review score
used as a metric when evaluating game design and development teams
\citep{Ribbens2012,Yu2021}. Games with low numbers of reviews may
be sold at a discount to encourage more reviews \citep{Kim2014}.
Trust in reviews is highest for voluntary reviews \citep{Petrosino2022}
as fake reviews are usually created for payment \citep{Bian2021,Angelis2021}.
Aggregation of reviews (e.g., via a systematic review) reduces bias
\citep{Bashir2019}.

However, game reviews can also be a way of communicating with the
design and development team \citep{Lin2018}, to effect change in
the game \citep{Fong2017,Yu2023}. Review processes that encourage
the player to think as a designer \citep{Kirschner2014} allow the
player to gain greater insight into their own understanding of the
game. Design insights are more valuable when the reviewer has greater
experience with the field and technology \citep{Faric2019}. Game
designers mine reviews to get insight into player preferences \citep{Sirbu2016,Urriza2021,kosmopoulos2020summarizing}.
Reviews also provide a support channel to report problems, request
features or ask questions \citep{Alfaro2021} of developers or other
players. Heuristic reviews are conducted within the development team
to test the design \citep{Livingston2010}.

Serious games, or those for educational purposes, have additional
goals and require reviews that assess how well they achieve these
goals \citep{Bas2019,Calderon2015,Shiratuddin2011} or suggest educational
improvements \citep{Alfaro2021}. The review is used to directly assess
the success of an educational game \citep{Giani2016}, particularly
if the structure of the review is specified in advance \citep{Petri2017}.
Standardizing the format of the review allows it to be used as an
assessment form \citep{Calderon2015}, and a literature review of
published reviews then helps identify trends and themes across these
reviews. Qualitative elements of the review provide insight into good
game design and game-based training principles that can be applied
in future game designs \citep{Coleman2014}.

Our own motivation: using reviews to identify relevant educational
resources is shared \citep{Caserman2020} where a standardized
serious game metadata format is employed to describe serious games so teachers/therapists
can find suitable games to use in class/therapy.

\subsubsection{Field}

The areas that $\exper$s target are summarized in Figure \ref{tab:Field-1}.
As may be expected based on the search terms used, most reviews relate
to games, with a smaller subset relating to virtual and augmented
reality. Despite the screening criteria mobile games and applications
are also represented as these tend to overlap with games and XR applications.

While most of these analyses of $\exper$s cover games in general
several do restrict their focus to particular areas within this field
in order to extract targeted insights. These foci cover particular
aspects of games such as genres (role playing, souls-like, multiplayer)
and purposes such as serious, educational, simulation, and therapy.
The few XR applications are all specialized, covering areas such as
games, social collaboration, health and education.

The reviews with a defined purpose cover areas of game design such
as play experience \citep{Kirschner2014,Zhu2016}, narrative \citep{Cho2020},
player expectations \citep{Straat2017}, addiction \citep{Balakrishnan2018}
and sexuality and gender \citep{Kohlburn2022}. Serious and educational
games are those with an educational focus and where entertainment
is a secondary outcome. Educational topics covered \citep{Bas2019,Calderon2015,Giani2016,Petri2017,Coleman2014,Caserman2020}
include cultural, professional, and social skills, and life decision
support training \citep{Calderon2015}, computing education \citep{Petri2017}
and ethics \citep{Cabellos2022}. Game reviews do also provide insight
into a wide range of other areas including information systems \citep{Bas2019,Zheng2021},
computer science \citep{Bas2019,Petri2017}, engineering \citep{Bas2019},
architecture \citep{Shiratuddin2011}, journalism \citep{Suominen2011},
economics \citep{Bas2019}, marketing \citep{Tsang2009,Zheng2021},
and health \citep{Calderon2015,Caserman2020,Bashir2019,Silva2022}.
Review ecosystems consisting of collections of reviews provide insight
into the domain of the reviews while also being relevant to many other
overlapping domains.

\begin{figure}
\begin{centering}
\begin{tabular}{|>{\raggedright}p{0.1\textwidth}|>{\raggedright}p{0.3\textwidth}|}
\hline 
{\footnotesize{}Application category} & {\footnotesize{}Sub-category}\tabularnewline
\hline 
\hline 
{\footnotesize{}games} & {\footnotesize{}any \citep{Eberhard2018,Kasper2019,Kirschner2014,Koehler2017,Lin2018,Livingston2011,Wang2020,Ribeiro2019,Viggiato2022,Fong2017,Urriza2021,Zhu2016,Baowaly2019,Ribbens2012,Li2021a,Zhu2015,Straat2017,Vieira2019,Wattanaburanon2016,Kohlburn2022,Cox2015,Kim2014,Soetedjo2022,Bian2021,Kwak2020,Cho2020,kosmopoulos2020summarizing,Busurkina2020,Santos2019,Ahn2017,Strt2017UsingUC,Meidl2021,Suominen2011,Livingston2010,Zagal2009,Bond2009,Zhu2010,Wang2021,JeongminSeo2023,Boric2023,Tong2021,SouzaGoncalves2020,Chand2022,Yu2023,Guzsvinecz2023,Philp2024},}{\footnotesize\par}

{\footnotesize{}souls-like \citep{Guzsvinecz2022}, PUBG \citep{Yu2021}}\tabularnewline
\hline 
 & {\footnotesize{}any mobile \citep{Ho2012,Lin2018,Petrovskaya2022},
mobile role playing \citep{Youm2022}}\tabularnewline
\hline 
 & {\footnotesize{}desktop \citep{Petrovskaya2022}}\tabularnewline
\hline 
 & {\footnotesize{}multiplayer \citep{Petrosino2022}}\tabularnewline
\hline 
 & {\footnotesize{}serious \citep{YanezGomez2016,Caserman2020,Calderon2015,Sirbu2016,Silva2022,alQallawi2021},
educational \citep{Giani2016,Petri2017,Coleman2014,Calderon2015,Balakrishnan2018,Cabellos2022}}\tabularnewline
\hline 
 & {\footnotesize{}simulation \citep{Bas2019}}\tabularnewline
\hline 
 & {\footnotesize{}therapy \citep{Phillips2021}}\tabularnewline
\hline 
{\footnotesize{}virtual reality} & {\footnotesize{}games \citep{Gao2022,Epp2021}}\tabularnewline
\hline 
 & {\footnotesize{}exergames \citep{Faric2019}}\tabularnewline
\hline 
 & {\footnotesize{}social and collaborative \citep{Deng2023}}\tabularnewline
\hline 
 & {\footnotesize{}relaxation and meditation \citep{Fagernas2021}}\tabularnewline
\hline 
{\footnotesize{}augmented reality} & {\footnotesize{}mobile games \citep{Zubair2021}}\tabularnewline
\hline 
 & {\footnotesize{}educational \citep{Alfaro2021}}\tabularnewline
\hline 
{\footnotesize{}mobile applications} & {\footnotesize{}\citep{Huang2015}}\tabularnewline
\hline 
{\footnotesize{}game-related settings} & {\footnotesize{}creative artifacts \citep{Balietti2016}, programming
\citep{WangReview2008}, game studies \citep{Barr2017}}\tabularnewline
\hline 
\end{tabular}
\par\end{centering}
\caption{\label{tab:Field-1}Field: application areas for $\exper$s}
\end{figure}

\subsection{\label{subsec:Reviewutilization-1}Review~utilization}

The next step in understanding what sustains a review ecosystem is
to ask who uses reviews, what they are used for and how they are used.

\subsubsection{Target~audience}

We make a loose distinction between customers and players although
these two categories overlap. Customers utilize reviews as recommendations
regarding financial decisions while players also use them to decide
how to invest time and effort to maximize rewards from game play \citep{Livingston2011}.
Player value includes benefits beyond just identifying quality of
the game. Reviews provide concepts, ideas and insights that enhance
the overall ability and sophistication of a game player. Reviews also
provide specific information about strategies and opportunities that
exist in the experience being reviewed, and that might have otherwise
been overlooked.

Similarly, reviewers (particularly amateur reviewers) are also members
of the player community. However, in their role as reviewers their
focus is on participating in the review ecosystem. Here they provide
feedback to their peers and use reviews to develop environments in
which they can explore their interests (playing games) and to develop
community standing. The target audience may be global, allowing reviewers
to have significant impact \citep{Wang2021}, or local where the reviewer
writes for their friends \citep{Boric2023} or those with similar
cultural backgrounds \citep{Tsang2009}.

The target audience identified in Figure \ref{tab:Audience-1} covers
diverse areas of the game production and deployment pipeline. At each
stage, a review provides a form of feedback. This provides insights
to designers and allows players to communicate with the developers
of their games. Game producers rely on reviews to inform players and
influence their purchasing decisions. Research academics producing
case studies or game analysis papers have similar motivations but
within a more restricted community. The review ecosystems that have
developed to support $\scoping$s can be self-sustaining, generating
new reviews even when no audience need has been identified \citep{Bashir2019}.

Educators and students are the other audience groups that are identified
in this analysis. Reviews satisfy their needs by providing a screening
mechanism to identify relevant game experiences and to validate their
quality.

\begin{figure}
\begin{centering}
\begin{tabular}{|>{\raggedright}m{0.08\textwidth}|>{\raggedright}m{0.08\textwidth}|>{\raggedright}p{0.24\textwidth}|}
\hline 
{\footnotesize{}Sector} & {\footnotesize{}Audience} & {\footnotesize{}Reason for using/creating review}\tabularnewline
\hline 
\hline 
\multirow{14}{0.08\textwidth}{{\footnotesize{}Game production}} & \multirow{2}{0.08\textwidth}{{\footnotesize{}Designers}} & {\footnotesize{}Ensuring assessment in simulation games is accurate
\citep{Bas2019}}\tabularnewline
\cline{3-3} 
 &  & {\footnotesize{}Understanding what makes an effective design \citep{Koehler2017,Shiratuddin2011,Wang2020,Straat2017,Soetedjo2022,Bond2009}}\tabularnewline
\cline{2-3} \cline{3-3} 
 & \multirow{3}{0.08\textwidth}{{\footnotesize{}Developers}} & {\footnotesize{}Understanding players and how to improve a game \citep{Lin2018,Coleman2014,Fong2017,Urriza2021,Guzsvinecz2023}}\tabularnewline
\cline{3-3} 
 &  & {\footnotesize{}As a measure of the success of a game \citep{Livingston2011,Giani2016,Petri2017}}\tabularnewline
\cline{3-3} 
 &  & {\footnotesize{}As a way to receive feedback from players \citep{Youm2022,Yu2021,Tong2021,alQallawi2021,WangReview2008}}\tabularnewline
\cline{2-3} \cline{3-3} 
 & {\footnotesize{}Customers} & {\footnotesize{}Making purchasing decisions \citep{Eberhard2018,Ho2012,Wang2020,Baowaly2019,Ribbens2012,Kohlburn2022,Cox2015,Kim2014,Livingston2010,Zagal2009,Zhu2010,Zheng2021}}\tabularnewline
\cline{2-3} \cline{3-3} 
 & {\footnotesize{}Publishers} & {\footnotesize{}To influence commercial outcomes (e.g., manage game
evolution, fake reviews) \citep{Bian2021,Angelis2021,Yu2023}}\tabularnewline
\cline{2-3} \cline{3-3} 
 & \multirow{4}{0.08\textwidth}{{\footnotesize{}Reviewers}} & {\footnotesize{}Contribute to a review ecosystem \citep{Kasper2019}}\tabularnewline
\cline{3-3} 
 &  & {\footnotesize{}Measuring the value of a game \citep{Koehler2017}}\tabularnewline
\cline{3-3} 
 &  & {\footnotesize{}Communicating with others in a player community \citep{Koehler2017,Tsang2009,Wattanaburanon2016,Petrosino2022,kosmopoulos2020summarizing,Suominen2011,Deng2023,Boric2023}}\tabularnewline
\cline{3-3} 
 &  & {\footnotesize{}Self-expression and developing personal standing in
the community \citep{Cho2020,Busurkina2020,Santos2019,Tong2021}}\tabularnewline
\cline{2-3} \cline{3-3} 
 & \multirow{3}{0.08\textwidth}{{\footnotesize{}Players}} & {\footnotesize{}Reflecting on own play to improve experience \citep{Kirschner2014,Straat2017}}\tabularnewline
\cline{3-3} 
 &  & {\footnotesize{}As a way to assess game quality prior to purchase
\citep{Livingston2011,Ribeiro2019,Sirbu2016,Urriza2021,Gao2022,Youm2022,Ribbens2012,Bond2009}}\tabularnewline
\cline{3-3} 
 &  & {\footnotesize{}To receive information and advice \citep{Kohlburn2022,Philp2024}}\tabularnewline
\hline 
\multirow{4}{0.08\textwidth}{{\footnotesize{}Education and training}} & \multirow{2}{0.08\textwidth}{{\footnotesize{}Teachers}} & {\footnotesize{}Using games with validated assessment \citep{Bas2019,Giani2016}}\tabularnewline
\cline{3-3} 
 &  & {\footnotesize{}To choose appropriate educational resources \citep{Petri2017,Caserman2020}}\tabularnewline
\cline{2-3} \cline{3-3} 
 & {\footnotesize{}Students} & {\footnotesize{}Using effective resources when learning \citep{Giani2016,YanezGomez2016}}\tabularnewline
\cline{2-3} \cline{3-3} 
 & {\footnotesize{}Peers} & {\footnotesize{}Peer feedback \citep{Barr2017,Friedrich2019}}\tabularnewline
\hline 
\multirow{2}{0.08\textwidth}{{\footnotesize{}Research}} & {\footnotesize{}Academics} & {\footnotesize{}Research into properties of games \citep{Bas2019,Calderon2015,Gao2022,Chand2022}}\tabularnewline
\cline{2-3} \cline{3-3} 
 & {\footnotesize{}Publishers} & {\footnotesize{}Process of sharing research \citep{BRYAND.2006,Garcia2017,Garcia2020}}\tabularnewline
\hline 
\end{tabular}
\par\end{centering}

\caption{\label{tab:Audience-1}Audience categories for $\exper$s}
\end{figure}

\subsubsection{Usage}

The Usage category is similar to Focus (section \ref{subsec:Focus-1}),
with the key difference being perspective: the review author provides
focus while the review consumer finds uses. Often these two perspectives
are not explicitly distinguished, with usage being focused around:
supporting purchase recommendations \citep{Eberhard2018,Ho2012,Kasper2019,Ribbens2012,Wattanaburanon2016,Kim2014,Bian2021,Angelis2021,Ahn2017,Bond2009},
providing insight to players \citep{Kirschner2014,Livingston2011,Faric2019},
rating and ranking games \citep{Koehler2017,Busurkina2020,SouzaGoncalves2020},
classifying and cataloging games \citep{Cho2020}, assessing and evaluating
games \citep{Bas2019,Calderon2015,YanezGomez2016}, providing feedback
to the game developers \citep{Lin2018,Petri2017,Coleman2014,Wang2021,Guzsvinecz2023,WangReview2008},
predicting game sales \citep{Tsang2009,Petrosino2022,Santos2019,Zhu2010,Tong2021},
and supporting the use of the game as an educational tool \citep{Giani2016,Petri2017,Caserman2020,Alfaro2021}.

Reviews can be critical in determining the success or failure of a
product \citep{Balietti2016,BRYAND.2006,Garcia2017}. They are preferred
over other sources of information because they are considered trustworthy
\citep{Bian2021,Zheng2021}, because they highlight value \citep{Guzsvinecz2022},
and because they provide objective information based on actual personal
experience \citep{Ribeiro2019,Li2021a,Zhu2015,Petrovskaya2022} rather
than marketing hype \citep{Youm2022,Kim2014}. Reviews reveal information
that may be missed during testing or user experience evaluation \citep{Straat2017}
such as aspects of the product most significant to users \citep{Vieira2019}
or how players reason about games \citep{Busurkina2020,Ahn2017,Silva2022,Deng2023}.
Poor reviews can influence perceptions, even for existing players
\citep{Philp2024}.

Collections of reviews are frequently data mined to extract insights
related to the experiences being reviewed. A range of game related
properties are derived through analysis and aggregation of reviews
such as: game quality \citep{Sirbu2016}, playability \citep{Zhu2015},
user sentiment \citep{Strt2017UsingUC}, treatment efficacy \citep{Bashir2019},
and developer performance \citep{Ribbens2012}. Review properties
can be a proxy for game metrics once correlations between the review
characteristics and the game measure are established \citep{Kohlburn2022}.
For example, longer reviews tend to correspond to lower levels of
satisfaction with the game \citep{Wang2020}, and some review properties
correlate with moral game themes \citep{Cabellos2022}. Reviews themselves
are used as a proxy for experiencing a game before purchase \citep{Busurkina2020,Huang2015,Wang2021}.
Rules for product design can be inferred from collections of reviews
\citep{Soetedjo2022}. Numerical scores based on review features support
efficient ranking and game comparisons \citep{kosmopoulos2020summarizing}.

Value is created when researchers analyze patterns of reviews to identify
insights that might emerge from the review \emph{ecosystem} \citep{Koehler2017}.
Review ecosystems provide an ideal research data source where a large
amount of data has already been collected, and without influence from
the research team \citep{Phillips2021,Yu2021,Chand2022}, and that
can provide insight into the community, its history and its conventions
\citep{Suominen2011,Zagal2009}. Reviews serve as a feedback mechanism
and can be regarded as a communication mechanism that ensures the
diverse perspectives of specialist groups \citep{Petri2017,Barr2017}
and collective opinions \citep{Li2021a,Wattanaburanon2016,Petrosino2022,alQallawi2021}
are shared to achieve consensus \citep{Huang2015}. Having specialist
stakeholders review a design is typical within an organization and
identifies issues prior to release \citep{Livingston2010}. The benefits
of review are implicit in a game review ecosystem where insights from
players and educators is fed back to developers to report bugs and
suggest improvements \citep{Viggiato2022,Fong2017,Urriza2021,Wattanaburanon2016,Alfaro2021}
and solutions \citep{Youm2022}. For eduXR experiences the reviews
reveal human interface issues, motion sickness and problems with interaction
affordances \citep{Gao2022}. Suggestion made within reviews are a
form of free customer support \citep{Baowaly2019} and product advertising
\citep{Angelis2021,Tong2021}. The reviews themselves can be used
as a training tool for developing review writing skills \citep{Coleman2014,Friedrich2019}
and as training in identifying the attributes to consider when evaluating
a system.

\subsubsection{Education}

A goal of this work is to investigate how game review ecosystem concepts
translate to eduXR. While education was not explicitly included in
the search criteria, education is still a strong theme in a number
of the sources identified. Only three of the sources identified use
the word ``education'' in their titles, although several others
are published in journals associated with this theme.

\begin{figure}
\begin{centering}
\includegraphics[width=0.4\paperwidth]{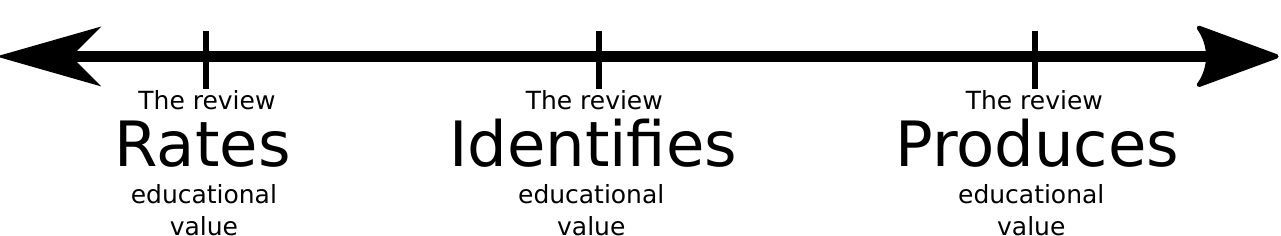}
\par\end{centering}
\caption{\label{fig:The-continuum-of-1}The continuum of educational purposes
of reviews.}
\end{figure}

The aspects of education in the reviews varies over the continuum
shown in Figure \ref{fig:The-continuum-of-1} ranging from measuring
educational properties at one end, to the review being part of the
educational process at the other end.
 These educational properties are used to make decisions about the
relevance of the overall experience (its ``fit for purpose'') \citep{Petri2017}
and whether it authentically conveys the intended skills, knowledge
\citep{Caserman2020} and assessment \citep{alQallawi2021}. Narrative
games implicitly provide educational value through their stories and
this can be revealed in reviews \citep{Cho2020}. Augmented reality
is a significant educational technology and reviews for these experiences
report on the associated educational area and level. However, such
educational branding can also be misused to target children as an
audience \citep{Alfaro2021}.

Further along the continuum is where the review reports on individual
educational qualities, such as realism, experiential learning through
play or skills such as decision making and team work \citep{Bas2019},
measures of student learning or quality of materials \citep{Giani2016},
game elements and their value \citep{Straat2017}, or learning objectives
and scaffolding \citep{Coleman2014}. Reporting on the game teaching
particular topics is central on the continuum. Here the focus is on
content such as software project management \citep{Calderon2015},
or where the review structure defines best practices in reporting
on game usability \citep{YanezGomez2016}.

The far side of the continuum is the educational value provided by
the review process itself. Reviews provide feedback to correct and
refine designs \citep{Shiratuddin2011,Zhu2016,Straat2017,WangReview2008,Friedrich2019}
and the reviewing process enhances understanding of topics such as
play \citep{Kirschner2014}. Reviews written by learners provide insight
into the educational value that they perceive \citep{Sirbu2016,Tsang2009,Barr2017}.
Students should learn to write reviews before learning to develop
the artifacts that would be reviewed \citep{Barr2017}. Filling in
a questionnaire to generate an $\exper$ is used to both evaluate
a learning experience and provides a mechanism for students to self-assess
the learning achieved \citep{Petri2017}.

\subsubsection{\label{subsec:Mechanics-1}Mechanics}

The processes for creating reviews can be reused when developing new
review ecosystems. The following themes emerged among the literature
reviewed:
\begin{itemize}
\item \textbf{Sources~of~information}: Reviews are created by players
based on their experience of playing the game \citep{Koehler2017,Wang2020},
as an assessment of the product \citep{Tsang2009}, or through an
analysis of the design \citep{Shiratuddin2011}. Typically only a
single experience is reviewed \citep{Zhu2016} leaving opportunities
for reviews that compare and contrast several experiences. The review
is an opinion that may represent the reviewer's mental model and preconceptions
of the game as much as the reality \citep{Tong2021}.

Conflict of interest and potential sources of bias would need to be
managed \citep{Coleman2014,Zhu2010}; for example if commercial interests
are involved when professional reviewers create the reviews \citep{Koehler2017,Angelis2021},
or when written by the developers as a way of marketing a game \citep{Coleman2014,Caserman2020}.
Reviews sites often host and distribute the same games that are being
reviewed \citep{Busurkina2020}. Blind reviewing is typically used
for peer review \citep{Sacco2020} while review sites can instead
disclose detailed reviewer metadata \citep{Philp2024}. Ecosystems
where issues identified in the review can be verified or discussed
are less likely to allow bias to propagate \citep{Shiratuddin2011}.
\item \textbf{Quality~control}: Reviews are assumed to be based on actual
player experience, with some reviews sites able to validate that the
reviewer does at least own a copy of the game \citep{kosmopoulos2020summarizing},
or has sufficient play experience \citep{Zhu2010}. However, it may
be possible for reviews to be created with different levels (including
zero) of play time \citep{Kasper2019,Guzsvinecz2023}, or be created
based on other reviews \citep{Livingston2011,YanezGomez2016}. Review
credibility can be established by providing information about the
reviewer \citep{Zheng2021}, such as their play and review history,
motivation for creating the review, or experience with game design
\citep{Tong2021}. Reviews created by professionals, e.g., journalists
are structured, rigorous and use appropriate terminology \citep{Ribbens2012,Cox2015,Suominen2011},
while amateur reviews are created spontaneously with no external incentives
\citep{Faric2019}. Reviewer interactions within a community can have
a detectable influence on review contents \citep{Tong2021}.

Review sites may prompt for particular information implying that unprompted
topics that occur frequently represent an intrinsic property of the
experience being reviewed \citep{Phillips2021}. Nostalgia leads to
positive reviews created long after game release \citep{Santos2019}.
Rigor is introduced by collecting and analyzing data according to
frameworks, scales and measures, defined processes and models with
standardized questions \citep{Giani2016}. These are then analyzed
and reported using descriptive statistics and hypothesis testing \citep{Petri2017}.
\item \textbf{Effort~involved}: Review sites can simplify the creation
of reviews by allowing short comments \citep{Koehler2017,Silva2022}
or even numerical ratings \citep{Ribeiro2019,Zheng2021} as a form
of review to quickly capture a holistic view of the game. Online reviews
are created and hosted on sites designed to support creation of reviews
\citep{Kim2014,Bian2021}. Negative reviews are often short but written
earlier \citep{Guzsvinecz2023}, with positive reviews requiring more
effort to describe the value in the experience, or to refute previous
reviews \citep{Kohlburn2022}. More complex processes exist that involve
recording and reviewing game play over many hours and where analysis
of this material drawn out through interviews and discussion with
others \citep{Kirschner2014,Koehler2017,YanezGomez2016}. Reviews
produced as published case studies follow formal research processes
such as interviews and focus groups, expert evaluations, surveys,
case studies, experiments, use of models and frameworks, pre- and
post-tests, heuristic evaluations or logs \citep{Koehler2017,Giani2016,YanezGomez2016}
with tools to support these processes \citep{Bashir2019}. In contrast
most casual reviews never mention the review process \citep{Zagal2009}.
Typically a review is created by a single author \citep{Balakrishnan2018}
although a review ecosystem is created by a community \citep{Meidl2021}.
\item \textbf{Stage}: Most reviews are created some time after playing the
game, particularly after the player is invested in the experience
\citep{Guzsvinecz2022}. There is value in creating the review while
playing \citep{YanezGomez2016} or reviewing while re-watching recordings
of the play experience \citep{Kirschner2014}. Experts typically create
reviews soon after a game release, while amateurs can still be creating
reviews years later \citep{Santos2019}. Priming by positive/negative
reviews before play has an impact on later reviews created \citep{Livingston2011}.
In a design context, reviews occur at stages of design (e.g., preliminary,
working drawings, schematic) \citep{Shiratuddin2011} or during testing
\citep{Livingston2010} so do not require a finished product before
a review can be completed. This theme suggests that context should
be considered in an $\exper$ ecosystem by recording the background
of the participants \citep{Wang2021,Zheng2021} and the structure
of the community that is established. Peer review systems include
a stage involving revision of the product after the review is received
\citep{Friedrich2019}.
\end{itemize}

\subsection{\label{subsec:Ecosystemmanagement-1}Ecosystem~management}

Collecting reviews involves creating, collating and managing these
assets on an ongoing basis. The interactions between multiple components
(the games being reviewed, the people doing the reviewing, and the
environment that hosts the reviews) is similar to an ecosystem resulting
from the interactions of organisms and their environment. Only a few
of the sources \citep{Koehler2017,Giani2016} explicitly identify
this overarching environment and the processes involved in managing
this although many of the others implicitly identify ecosystem structures
and mechanisms.

\subsubsection{Structure}

The components of the review ecosystem include the groups of stakeholders
that are involved in creating, curating and benefiting from the reviews
that are produced. Figure \ref{tab:Structure-1} enumerates the groups
that are identified. The names used for each class of stakeholder
represent the terms used in each cited source that interprets their
role and function.

Four categories of ecosystem elements are identified. These categories
are consistent with social communication theory that identifies: the
review writers (communicators), the review users (communicatees),
and the review hosts (channels). The fourth category of review modifiers
matches the communication category of response; the action taken after
reading the review but does also include modifications to the review
during any stage before and after publication.

Review writers produce reviews. These are the game players although
various special interest groups can also be identified. From a commercial
perspective, professional review writers and the game developers have
a financial interest in the content of the review. Educators who intend
to use the game in class will evaluate its suitability to meet learning
outcomes. Academic analysis of particular games results in papers
from academic reviewers. A less common source of reviews are those
created for quality control where the review is used internally to
refine development of the game. The subtle distinction between game
owner (purchaser) and game player as a source of reviews relates both
to a form of validation (that the reviewer has actually played the
game before writing the review) and a source of bias (the owner has
already invested in the game).

Review users gain the most direct benefit from the review ecosystem.
These are game consumers for whom the review provides guidance on
game choice. Consumers are players viewed from the perspective of
purchasing games and consuming reviews. Game designers and developers
are also well represented since reviews can determine the success
of a game and the insights communicated within the review provide
guidance on refining the game. Review insights are valuable as reviewers
may have experience with a wider range of games than many game developers
\citep{Zagal2009}. Reviews identify games that can be repurposed
for education, although relatively few sources explicitly mention
the need to verify that the information provided by a game is accurate
\citep{Caserman2020}. Internal reviews are another form of the review
ecosystem that operates within organizations, to the overall benefit
of the organization.

The ecosystem extends beyond the consumption of raw reviews. Collections
of reviews are used for text mining. Review rating systems provide
recommendations or can be used to fit predictive models. Assumptions
around the nature of reviews can also be challenged; for example that
a review of a game is independent of the player. Each player has a
unique background which defines their play experience and has their
own personal approach to interacting with the game.

The environment in which reviews exist is the fourth category. Game
reviews exist in online game review sites, but are also hosted in
publications created for this purpose. Unlike physical publications,
online reviews exist in a space without boundaries \citep{Zhu2010}.
These sites may supplement reviews with information such as the reviewer
history and reputation \citep{Wang2021}. The reviews support their
hosting sites by directing purchase decisions for the products reviewed,
but can add value by providing the basis for a social community that
provides trusted, word-of-mouth recommendations. Other forms of review,
such as academic papers reporting case studies, exist in paper repositories
and databases.

\begin{figure*}
\begin{centering}
\begin{tabular}{|>{\raggedright}m{0.1\textwidth}|>{\raggedright}p{0.15\textwidth}|>{\raggedright}p{0.2\textwidth}|>{\raggedright}p{0.15\textwidth}|}
\hline 
{\footnotesize{}Ecosystem elements} & {\footnotesize{}Stakeholder group} & {\footnotesize{}Purpose} & {\footnotesize{}Identified in}\tabularnewline
\hline 
\hline 
\multirow{8}{0.1\textwidth}{{\footnotesize{}Review writers}} & {\footnotesize{}Researchers / Academics} & {\footnotesize{}Evaluating games and reasoning about their properties} & {\footnotesize{}\citep{Bas2019,Kirschner2014,Koehler2017,Wang2020,BRYAND.2006}}\tabularnewline
\cline{2-4} \cline{3-4} \cline{4-4} 
 & {\footnotesize{}Game players / Users} & {\footnotesize{}Sharing opinions} & {\footnotesize{}\citep{Eberhard2018,Kasper2019,Kirschner2014,Koehler2017,Lin2018,Livingston2011,Giani2016,YanezGomez2016,Ribeiro2019,Urriza2021,Gao2022,Youm2022,Balakrishnan2018,Li2021a,Zhu2015,Straat2017,Vieira2019,Wattanaburanon2016,Guzsvinecz2022,Cox2015,Petrovskaya2022,Kim2014,Zubair2021,kosmopoulos2020summarizing,Busurkina2020,Santos2019,Strt2017UsingUC,Deng2023}}\tabularnewline
\cline{2-4} \cline{3-4} \cline{4-4} 
 & {\footnotesize{}Game Owners} & {\footnotesize{}Rating of their property} & {\footnotesize{}\citep{Lin2018,Wang2020}}\tabularnewline
\cline{2-4} \cline{3-4} \cline{4-4} 
 & {\footnotesize{}Professional / Export review writers} & {\footnotesize{}Make recommendations and critical assessment} & {\footnotesize{}\citep{Kasper2019,Koehler2017,Livingston2011,YanezGomez2016,Coleman2014,Viggiato2022,Ribbens2012,Straat2017,Wattanaburanon2016,Cox2015,Santos2019,SouzaGoncalves2020}}\tabularnewline
\cline{2-4} \cline{3-4} \cline{4-4} 
 & {\footnotesize{}Game Creators / Developers} & {\footnotesize{}Providing information on their product} & {\footnotesize{}\citep{Petri2017}}\tabularnewline
\cline{2-4} \cline{3-4} \cline{4-4} 
 & {\footnotesize{}Reviewers / Testers} & {\footnotesize{}Performing quality control, contributing to game development} & {\footnotesize{}\citep{Shiratuddin2011,Coleman2014,Fong2017,Faric2019,Vieira2019,Tsang2009,Livingston2010,Zagal2009,Bashir2019,Zheng2021,Tong2021}}\tabularnewline
\cline{2-4} \cline{3-4} \cline{4-4} 
 & {\footnotesize{}Educators} & {\footnotesize{}Reporting educational content} & {\footnotesize{}\citep{Coleman2014}}\tabularnewline
\cline{2-4} \cline{3-4} \cline{4-4} 
 & {\footnotesize{}Students} & {\footnotesize{}Creating reviews while learning} & {\footnotesize{}\citep{Sirbu2016}}\tabularnewline
\hline 
\multirow{8}{0.1\textwidth}{{\footnotesize{}Review users}} & {\footnotesize{}Educators} & {\footnotesize{}Selecting games for use in class} & {\footnotesize{}\citep{Bas2019,Giani2016,Petri2017,Coleman2014,Caserman2020}}\tabularnewline
\cline{2-4} \cline{3-4} \cline{4-4} 
 & {\footnotesize{}Students} & {\footnotesize{}Selecting games to support education} & {\footnotesize{}\citep{Giani2016,Petri2017}}\tabularnewline
\cline{2-4} \cline{3-4} \cline{4-4} 
 & {\footnotesize{}Domain experts} & {\footnotesize{}Validating material presented} & {\footnotesize{}\citep{Caserman2020}}\tabularnewline
\cline{2-4} \cline{3-4} \cline{4-4} 
 & {\footnotesize{}Researchers / Academics} & {\footnotesize{}Using reviews as a source of data} & {\footnotesize{}\citep{Gao2022,Alfaro2021,Busurkina2020,Suominen2011}}\tabularnewline
\cline{2-4} \cline{3-4} \cline{4-4} 
 & {\footnotesize{}Customers / Consumers} & {\footnotesize{}Making purchase decisions through word of mouth recommendations,
of products from review sites} & {\footnotesize{}\citep{Eberhard2018,Ho2012,Kasper2019,Lin2018,Wang2020,Coleman2014,Urriza2021,Baowaly2019,Ribbens2012,Tsang2009,Wattanaburanon2016,Kim2014,Petrosino2022,Yu2021,Angelis2021,Santos2019,Strt2017UsingUC,Suominen2011,Wang2021,Deng2023}}\tabularnewline
\cline{2-4} \cline{3-4} \cline{4-4} 
 & {\footnotesize{}Game developers / producers / designers / writers} & {\footnotesize{}Feedback on games, to support refinement} & {\footnotesize{}\citep{Eberhard2018,Kasper2019,Kirschner2014,Lin2018,Livingston2011,Petri2017,Shiratuddin2011,Coleman2014,Viggiato2022,Fong2017,Youm2022,Zhu2016,Ribbens2012,Zhu2015,Straat2017,Vieira2019,Tsang2009,Wattanaburanon2016,Petrovskaya2022,Alfaro2021,Soetedjo2022,Petrosino2022,Zubair2021,kosmopoulos2020summarizing,Busurkina2020,Santos2019,Strt2017UsingUC,Suominen2011,Zagal2009,Tong2021,WangReview2008}}\tabularnewline
\cline{2-4} \cline{3-4} \cline{4-4} 
 & {\footnotesize{}Owner / Publisher (of the game) / Marketer / Editor} & {\footnotesize{}Ensuring product is of an appropriate standard, and
commercially successful} & {\footnotesize{}\citep{Shiratuddin2011,Zhu2015,Tsang2009,Wattanaburanon2016,Kim2014,Petrosino2022,Yu2021,Angelis2021,Yu2023,BRYAND.2006,Barr2017,Garcia2017}}\tabularnewline
\cline{2-4} \cline{3-4} \cline{4-4} 
 & {\footnotesize{}Regulators} & {\footnotesize{}Third parties who need insight into game properties} & {\footnotesize{}\citep{Petrovskaya2022}}\tabularnewline
\hline 
\multirow{3}{0.1\textwidth}{{\footnotesize{}Review modifiers}} & {\footnotesize{}Raters of reviews} & {\footnotesize{}Identify most relevant reviews} & {\footnotesize{}\citep{Eberhard2018}}\tabularnewline
\cline{2-4} \cline{3-4} \cline{4-4} 
 & {\footnotesize{}Funders and editors} & {\footnotesize{}Determine which reviews are created and published
(in academic journals)} & {\footnotesize{}\citep{Bashir2019}}\tabularnewline
\cline{2-4} \cline{3-4} \cline{4-4} 
 & {\footnotesize{}Game} & {\footnotesize{}Review is an interpretation of play, in an interactive
game} & {\footnotesize{}\citep{Kirschner2014}}\tabularnewline
\hline 
\multirow{4}{0.1\textwidth}{{\footnotesize{}Review hosts}} & {\footnotesize{}Academic paper repository} & {\footnotesize{}Dissemination of academic research} & {\footnotesize{}\citep{Bas2019,YanezGomez2016,Bashir2019}}\tabularnewline
\cline{2-4} \cline{3-4} \cline{4-4} 
 & {\footnotesize{}Book / Magazines / Publishers (of reviews)} & {\footnotesize{}Curated collection of structured reviews} & {\footnotesize{}\citep{Coleman2014,Tsang2009}}\tabularnewline
\cline{2-4} \cline{3-4} \cline{4-4} 
 & {\footnotesize{}Commercial site selling games} & {\footnotesize{}Reviews support purchasing decisions} & {\footnotesize{}\citep{Eberhard2018,Ho2012,Lin2018,Wang2020,Petrosino2022}}\tabularnewline
\cline{2-4} \cline{3-4} \cline{4-4} 
 & {\footnotesize{}Commercial site hosting reviews} & {\footnotesize{}Reviews and social ecosystem drive traffic to site} & {\footnotesize{}\citep{Kasper2019,Angelis2021,Zhu2010,Huang2015,Zheng2021}}\tabularnewline
\hline 
\end{tabular}
\par\end{centering}
\caption{\label{tab:Structure-1}Structure: components of a review ecosystem}
\end{figure*}

\subsubsection{\label{subsec:Incentives-1}Incentives}

Effort is required to produce quality reviews \citep{Bianchi2018}.
Games utilize a wide range of incentive mechanisms, both within the
game to encourage the player to follow rules and achieve goals and
externally to incentivise the purchase of the game and game related
content. Review ecosystems, particularly those that support access
to games and to game play, employ incentives to encourage participation
and contributions.

The incentive mechanisms reported are relatively mundane. For academic
papers the incentive mechanisms are inherited from those for publishing
research \citep{Bas2019}. Peer review communities share both the
effort and benefits of reviewing \citep{Barr2017}. Such academic
peer review systems contain multiple roles and many feedback loops
\citep{Balietti2016}. There is little incentive to standardize the
$\scoping$ process \citep{Giani2016} (while noting that the papers
themselves do follow publisher guidelines). New $\scoping$s are only
created once the field has developed sufficiently to justify an update
\citep{Bashir2019}. Researchers also create games for particular
purposes \citep{YanezGomez2016} which provides an incentive for producing
reviews in the form of case studies. Free metadata useful for research
that is provided through review ecosystems would be costly to produce
via other mechanisms \citep{Cho2020}. Analysis of existing review
systems identifies structures that can be formalized to guide future
review writers \citep{Bond2009}.

Paid experts are more prolific in producing reviews \citep{Santos2019}.
Extrinsic motivation applies when designated experts are commissioned
to produce reviews \citep{Coleman2014,Ribbens2012,Tsang2009,Zheng2021}
or where review processes are managed within a team \citep{Shiratuddin2011}.
Payment for reviews is an explicit incentive mechanism \citep{Lin2018}
but is reflected in that longer reviews are created. Similarly, those
receiving professional benefits from reviews (e.g., educators choosing
appropriate tools to support their work) are implicitly encouraged
to utilize sources of informative reviews and to support such a system
\citep{Calderon2015}. Some groups of players are not able to directly
produce their own reviews (e.g., young children) and so their perspective
needs to be captured indirectly by others who observe or interpret
\citep{YanezGomez2016}. Pandemics produce motivation in the form
of need to play combined with additional free time to provide an opportunity
to create reviews \citep{Petrosino2022}.

Community game review sites are a form of emergent ecosystem \citep{Suominen2011},
each with their own style and conventions for reviews. Emergent systems
\citep{dewolf:definition:emergence} benefit from large numbers of
contributors and low barriers to entry as this allows messages from
individuals to reach a wide audience \citep{Petrosino2022}. Comment
sharing facilities are provided with no further incentive mechanism
\citep{Lin2018} beyond providing a way to share word-of-mouth opinions
\citep{Ho2012,Tong2021} or as a way to influence game developers
\citep{Lin2018} offers the benefit of addressing issues affecting
enjoyment of the game. Social standing is enhanced by reviewing popular
games \citep{Lin2018,kosmopoulos2020summarizing}. Community ranking
is an incentive as evidenced by greater levels of reviewer engagement
with small independent game developers \citep{Lin2018}. Commenting
directly on an experience ensures that the opinions expressed are
directly related to that experience \citep{Silva2022}. Anonymous
reviews remove constraints on what can be expressed which is both
a pro and a con \citep{Kohlburn2022,Sacco2020}. Conversely the number
of friends listed for an identified reviewer \citep{Zheng2021} and
privacy of play history \citep{Philp2024} affects the perception
of the review . Participating in a review ecosystem changes the way
players assess games and develops increasing sophistication in reasoning
about them \citep{Straat2017}. Setting an expectation that reviews
can be short \citep{Koehler2017} produces a greater number and variety
of reviews that can be distilled to reveal the wisdom of the crowd.
Scarcity mechanisms (limit of one review per game) provide incentives
for quality reviews \citep{Lin2018}. Recommendation systems need
to be seeded with some user created reviews \citep{Meidl2021} requiring
a contribution before benefiting from the ecosystem.

Trust, accuracy and consensus increases with the number of reviews
\citep{Kim2014,Huang2015,BRYAND.2006} and is eroded by fake reviews
\citep{Angelis2021,Tong2021} and commercial conflicts of interest
\citep{Boric2023}. A helpfulness rating assigned to reviews is a
meta-review, where the incentive to provide these would be linked
to how these support purchasing decisions. The calculations used to
produce such scores need to be transparent to avoid the perception
of bias \citep{Boric2023}. Scoring reviews through a helpfulness
rating \citep{Eberhard2018} helps increase the ranking and visibility
of the review \citep{Baowaly2019,Wang2021} but can discriminate against
both unpopular and new products that have few reviews \citep{Wang2021}.
Helpfulness ratings can degenerate into being measures of consensus
rather than a quality rating for a review \citep{Kasper2019}. Early
reviews can shape the tone of later ones \citep{Livingston2011} hence
there is an incentive to ensure accurate (or positive) early feedback.

In cases where developers provide information about their own games,
there is an incentive to shape perceptions of the games through this
\citep{Petri2017}. Developers benefit from a review ecosystem by
gaining an understanding of reasoning processes used by players, and
insights into the strengths and weakness of other games \citep{Viggiato2022,Livingston2010,Tong2021}.
The ecosystem reveals a consensus \citep{Urriza2021} that can be
considered to represent an objective viewpoint \citep{Youm2022}.
The commercial benefits of a review ecosystem extend beyond just promoting
sales and extend to community building within many of the online review
sites being hosted on the same platforms that sell the games \citep{Ribbens2012}.

\subsubsection{Value}

In terms of game theory different stakeholders are incentivised to
produce reviews because the strategies leading to review creation
have a higher payoff than other strategies. This payoff, the difference
between reward and cost, can be financial \citep{Lin2018,YanezGomez2016,Garcia2020}
but is mostly provided as other forms of benefit.

Figure \ref{tab:Value-1} summarizes the various rewards and costs
that have been identified in review ecosystems. Rewards concentrate
on the benefits of creating reviews, as opposed to additional value
being associated with the products reviewed. Review authors receive
direct benefits that can be financial, particularly for professional
or expert reviewers, or altruistic in that reviews increase the quality
of the games they play, or intangible such as higher standing in the
gaming community when quality ratings are applied to reviews. Despite
these rewards, it is possible that reviewers are intrinsically motivated
and would create reviews anyway \citep{Faric2019,Garcia2017}. Social
media effectively provides unlimited free reviews \citep{Silva2022}.
This is balanced against the downsides: writing reviews requires effort.
In some cases, there is particular disincentive (perceived lack of
novelty) for writing reviews for already reviewed products, both as
game reviews but also academic studies \citep{Petri2017}, despite
the benefits that reproducibility would provide.

Other stakeholders are motivated to manipulate incentives in a review
ecosystem. Developers benefit from the feedback provided to the development
process. Reviews identify good design practices enabling better quality
products. They help connect products with the people who can best
utilize them. The trade-off is the effort involved in monitoring a
continuous stream of reviews resulting from a thriving ecosystem,
and the risk that reviews may be subverted to promote other agendas
(e.g., review bombing \citep{Kasper2019}). The gaming community uses
the review ecosystem to spend money efficiently but also to identify
experiences that add value to the participant. The number of reviews
represent public exposure that is often as significant to publishers
as the review content \citep{Cox2015}.

A review ecosystem functions as a collaborative social network \citep{Shiratuddin2011}.
The costs to the gaming community is the effort involved in moderating
reviews, usually through review rating, but also the risk that early
poor reviews can influence the tone of later reviews \citep{Livingston2011}
which can introduce bias to the community. Educators and trainers
receive a similar benefit to developers when the relevant experiences
can be identified and used for their intended purpose or adapted.
Reviews are a way to measure the quality of the educational experience.
Quality assessment is also the target of reviews in the form of academic
case studies. The financial impact of reviews provides both rewards
and costs when they influence product sales positively or negatively.

The academic community represents an environment with a long established
review ecosystem (in the form of published papers). While full details
of the incentive system used in academic publishing is beyond the
scope of the sources reviewed the principles involved are worth considering
when developing any review ecosystem.

\begin{figure*}
\begin{centering}
\begin{tabular}{|>{\raggedright}p{0.1\textwidth}|>{\raggedright}p{0.3\textwidth}|>{\raggedright}p{0.3\textwidth}|}
\hline 
{\footnotesize{}Context} & {\footnotesize{}Reward} & {\footnotesize{}Cost}\tabularnewline
\hline 
\hline 
{\footnotesize{}Reviewer} & {\footnotesize{}Review scoring (ranking, helpfulness rating) \citep{Kasper2019}.}{\footnotesize\par}

{\footnotesize{}Recognition as a reviewer \citep{Eberhard2018}, social
standing \citep{Lin2018}, altruism \citep{Kohlburn2022}.}{\footnotesize\par}

{\footnotesize{}An opportunity to express yourself \citep{Guzsvinecz2022}.}{\footnotesize\par}

{\footnotesize{}Creating better game play \citep{Kirschner2014,Wang2020,Phillips2021,Tong2021}.}{\footnotesize\par}

{\footnotesize{}Payment \citep{Lin2018,YanezGomez2016}.} & {\footnotesize{}Effort (time and resources) \citep{Eberhard2018,Lin2018,Shiratuddin2011,Cho2020}.}{\footnotesize\par}

{\footnotesize{}Replicated reviews reduce reward \citep{Petri2017,Bashir2019}.}{\footnotesize\par}

{\footnotesize{}Scores as rewards incentivize subverting these mechanisms
with biased reviews \citep{Baowaly2019,Straat2017}, adding cost of
screening and verifying reviews \citep{Bian2021,Boric2023}.}\tabularnewline
\hline 
{\footnotesize{}Product development} & {\footnotesize{}A standard for evaluating new products \citep{Calderon2015}.}{\footnotesize\par}

{\footnotesize{}Identify best practices in design \citep{Calderon2015,Wang2020,Urriza2021,Wang2021}.}{\footnotesize\par}

{\footnotesize{}Provide relevant and specific information to developers
\citep{Petri2017,Fong2017,Gao2022,Wattanaburanon2016,kosmopoulos2020summarizing},
and to identify the cause of issues \citep{Soetedjo2022}.}{\footnotesize\par}

{\footnotesize{}Problems fixed in early reviews cost less \citep{Shiratuddin2011}.}{\footnotesize\par}

{\footnotesize{}Ensure products are used for their intended purpose
\citep{Caserman2020}.} & {\footnotesize{}Review bombing is a risk \citep{Kasper2019,Epp2021}.}{\footnotesize\par}

{\footnotesize{}Large numbers of reviews become time consuming to
monitor and extract information from \citep{Lin2018,Wattanaburanon2016}.}{\footnotesize\par}

{\footnotesize{}Emotional elements need to be removed to find objective
recommendations \citep{Santos2019}.}\tabularnewline
\hline 
{\footnotesize{}Gaming community} & {\footnotesize{}The gaming experience starts with finding and assessing
games via reviews (play is only part of the experience) \citep{Straat2017}.}{\footnotesize\par}

{\footnotesize{}Communities for social games overlap with review communities
\citep{Petrosino2022}.}{\footnotesize\par}

{\footnotesize{}Value for money when purchasing games \citep{Eberhard2018,Petrosino2022}.}{\footnotesize\par}

{\footnotesize{}Protect against low quality products \citep{Youm2022}.}{\footnotesize\par}

{\footnotesize{}Trusted, and positive reviews have greater value \citep{Ho2012,Huang2015},
provide insight into products \citep{kosmopoulos2020summarizing}.}{\footnotesize\par}

{\footnotesize{}Reviews as a form of community collaboration \citep{Shiratuddin2011,Zheng2021},
with debate via review increasing the value \citep{Kim2014}. Online
reviews are available at any time \citep{Zhu2010}.}{\footnotesize\par}

{\footnotesize{}Review ecosystems ensure issues covered represent
the wider community \citep{Zhu2016}.} & {\footnotesize{}Effort involved in rating reviews \citep{Eberhard2018,Tong2021}.
Ratings are required to identify relevant reviews \citep{Tsang2009,Bashir2019}.}{\footnotesize\par}

{\footnotesize{}Negative reviews influence later reviews \citep{Livingston2011},
fake reviews erode trust \citep{Bian2021}.}{\footnotesize\par}

{\footnotesize{}Community values differ requiring review standards
adapt when sharing between communities \citep{Tsang2009}.}{\footnotesize\par}

{\footnotesize{}Expert curation may be required \citep{Santos2019,Suominen2011}.}{\footnotesize\par}

{\footnotesize{}Negative reviews make existing players appear incompetent
\citep{Philp2024}.}\tabularnewline
\hline 
{\footnotesize{}Educational and serious game community} & {\footnotesize{}Identify applications to support training \citep{Giani2016,Petri2017,Caserman2020},
reducing cost of ensuring fit-for-purpose \citep{Phillips2021}. Measure
of educational impact \citep{Petri2017}. Improved learning \citep{WangReview2008,Barr2017}.
Community participation \citep{Barr2017}.} & {\footnotesize{}Presenting information accurately can require expert
review \citep{Caserman2020}, or personal experience with the product
\citep{Cho2020}. Poor quality submissions create extra effort \citep{WangReview2008}.}\tabularnewline
\hline 
{\footnotesize{}Commercial interests} & {\footnotesize{}Positive (and fake \citep{Angelis2021}) reviews drive
purchases and provide commercial reward \citep{Ho2012,Ribeiro2019,Ribbens2012,Cox2015,Petrosino2022,Huang2015},
including as word-of-mouth recommendations \citep{Kim2014,Zhu2010}
and for new products without history \citep{Livingston2010}.}{\footnotesize\par}

{\footnotesize{}Reviews provide free support for products \citep{Ribeiro2019},
including details of platform specific features for platform spanning
games \citep{Petrovskaya2022}.}{\footnotesize\par}

{\footnotesize{}Review sites can sell advertising \citep{Ribbens2012},
support product sales \citep{Kim2014} and seed review ecosystems
\citep{kosmopoulos2020summarizing,Zheng2021}.} & {\footnotesize{}Negative reviews have financial costs \citep{Ho2012}.}{\footnotesize\par}

{\footnotesize{}Cost of creating a review site, including costs of
paid reviewers \citep{Suominen2011}.}\tabularnewline
\hline 
{\footnotesize{}Academic research} & {\footnotesize{}Reviews can be academic papers \citep{Bas2019}.}{\footnotesize\par}

{\footnotesize{}The research value of review ecosystems increases
with the number of reviews \citep{Zubair2021}. Reviews are an alternative
to evaluating a product directly \citep{Zubair2021}. Competition
encourages innovation \citep{Balietti2016}. Reviewer recognition
and financial rewards \citep{Bianchi2018,Garcia2017,Garcia2020}.} & {\footnotesize{}Reviews are not created to support research so not
directly useful in their original form \citep{Faric2019}. Very large
qualitative data sets can be a challenge \citep{Wang2021}. Competition
produces bias \citep{Balietti2016}. Effort required \citep{Garcia2017}.}\tabularnewline
\hline 
\end{tabular}
\par\end{centering}
\caption{\label{tab:Value-1}Value, in terms of reward and cost associated
with a review ecosystem}
\end{figure*}

\subsubsection{Quality}

Reviews cover a wide range of topics confounding comparison of specific
elements of a game \citep{Faric2019,Busurkina2020,Zheng2021}. A measure
to compare reviews or assess qualities of a review is required to
prevent the ecosystem being saturated with large numbers of poor quality
reviews. The goal of a review ecosystem should be to link user attitudes
through the reviews to the issue causing the attitudes \citep{Strt2017UsingUC}.
The different approaches to assessing this quality in reviews are
listed in Figure \ref{tab:Quality-1}. Care is taken to focus on comparison
mechanisms for $\exper$s, with many sources also discussing quality
measures for the game being reviewed \citep{Zhu2010}. These overlap
with the quality of the review being linked to the way in which it
describes particular game attributes such as challenge, conflict,
interaction, immersion, narrative and game rules and goals. Other
measures of review quality relate to the structure of the review itself,
such as the word count, readability and sentiments expressed. Reviews
need to describe the same version of an experience when comparing
qualities \citep{Wattanaburanon2016} and avoid introducing external
factors such as the reputation of the publisher \citep{Guzsvinecz2022}.

Approaches to identifying review quality range from manual approaches,
where ratings are assigned by the users of the platform hosting the
reviews, to automated strategies based on forms of text analysis.
Criteria, motivation, culture and standards differ between groups
of players which complicates the use of numerical scores for comparisons
\citep{Gao2022,Faric2019,Tsang2009,Petrovskaya2022,Santos2019,Epp2021,Boric2023}.
Aggregation encourages homogeneity and can disadvantage novel products
\citep{BRYAND.2006}. Automated strategies identify large numbers
of potential features that can exist in a review, and identify a relevant
subset of these features by correlating their presence to manually
assigned quality measures \citep{Vieira2019}. Trends and insights
from the most highly rated reviews efficiently summarize larger sets
of reviews and provide value to those utilizing the review ecosystem
\citep{Wang2021}. The diversity in the metrics used (despite some
commonalities) indicates that an ideal quality measure for a review
ecosystem has not yet been established.

\begin{figure}
\begin{centering}
\begin{tabular}{|>{\raggedright}p{0.12\textwidth}|>{\raggedright}p{0.3\textwidth}|}
\hline 
{\footnotesize{}Metric} & {\footnotesize{}How it is measured}\tabularnewline
\hline 
\hline 
{\footnotesize{}Quality} & {\footnotesize{}Based on criteria such as: details of evaluation method,
and features assessed \citep{Calderon2015}, standard of writing \citep{Ribbens2012}.}\tabularnewline
\hline 
{\footnotesize{}Quantity} & {\footnotesize{}Subjective variations in individual reviews are reduced
by aggregating large numbers of reviews \citep{Kim2014,Kwak2020,Wang2021,Tong2021}.}\tabularnewline
\hline 
{\footnotesize{}Helpfulness} & {\footnotesize{}User ratings of reviews, also fitted model based on
review similarity, structure, readability, and text content features
\citep{Eberhard2018,Baowaly2019,Kasper2019}.}\tabularnewline
\hline 
{\footnotesize{}Taxonomy fit \citep{Koehler2017}} & {\footnotesize{}Validating taxonomy of reviews \citep{Bedwell2012},
including features of: adaptation, assessment, challenge, conflict,
control, fantasy, interaction, rules and goals.}\tabularnewline
\hline 
{\footnotesize{}Scores} & {\footnotesize{}Using review measures \citep{Balietti2016}, distributions
and histograms \citep{Lin2018,kosmopoulos2020summarizing}, weighted
aggregations \citep{Straat2017,Livingston2010}, probability estimation
\citep{Garcia2017}.}\tabularnewline
\hline 
{\footnotesize{}Bias} & {\footnotesize{}Correlations in ratings between reviews \citep{Livingston2011}
identifies issues, e.g., fake reviews \citep{Bian2021,Bashir2019}.}\tabularnewline
\hline 
{\footnotesize{}Sales} & {\footnotesize{}Professional reviews measure product and affect sales,
consumer reviews can be caused by sales \citep{Cox2015}.}\tabularnewline
\hline 
{\footnotesize{}Validity} & {\footnotesize{}Instruments (reviews) should have: applicability,
utility, validity and reliability \citep{Giani2016}, provided in
good faith by players \citep{Silva2022} or experts \citep{Zheng2021}.}\tabularnewline
\hline 
{\footnotesize{}Predictive ability} & {\footnotesize{}Reviews must be able to predict: user satisfaction
\citep{Wang2020}, emotion and engagement \citep{Sirbu2016}.}\tabularnewline
\hline 
{\footnotesize{}Reputation} & {\footnotesize{}Review is trusted in proportion to perceived agreement
with previous reviews \citep{Ribeiro2019,Meidl2021,Tong2021}.}\tabularnewline
\hline 
{\footnotesize{}Feature extraction} & {\footnotesize{}Reviews are suited to feature extraction (e.g., natural
language processing \citep{Zhu2016,Fagernas2021,Alfaro2021,alQallawi2021,Guzsvinecz2023})
to extract: advantages, comparisons \citep{Viggiato2022}, summaries
\citep{Urriza2021}, game specific information \citep{Cho2020}, sentiment
\citep{Yu2023}, topics \citep{Deng2023} or keywords \citep{JeongminSeo2023}.}\tabularnewline
\hline 
{\footnotesize{}Ranking} & {\footnotesize{}The relative ranking of two games can be established
using reviews \citep{Fong2017,Zagal2009}.}\tabularnewline
\hline 
{\footnotesize{}Trends} & {\footnotesize{}Rather than ranking, metrics identify trends associated
with particular games \citep{Youm2022,Petrosino2022}, clusters of
reviews \citep{Ahn2017} or review style \citep{Suominen2011}.}\tabularnewline
\hline 
\end{tabular}
\par\end{centering}

\caption{\label{tab:Quality-1}Quality measures for $\exper$s}
\end{figure}

\subsubsection{Environment}

A review ecosystem exists in a particular environment. Reviews flourish
in game distribution platforms which also provide the opportunity
to comment or review individual products \citep{Livingston2011} and
can expose play habits as part of the review \citep{Philp2024}. Examples
of these include Steam \citep{Eberhard2018,Lin2018,Viggiato2022,Urriza2021,Baowaly2019,Li2021a,Vieira2019,Wattanaburanon2016,Kohlburn2022,Guzsvinecz2022,Petrovskaya2022,Soetedjo2022,Petrosino2022,Phillips2021,Bian2021,Yu2021,Cho2020,kosmopoulos2020summarizing,Busurkina2020,Ahn2017,Wang2021,Epp2021,Deng2023,Tong2021,Yu2023,Guzsvinecz2023,Philp2024},
Apple App Store \citep{Ho2012,alQallawi2021}, Google Play Store \citep{Youm2022,Balakrishnan2018,Petrovskaya2022,Alfaro2021,Zubair2021,alQallawi2021,Chand2022},
Amazon \citep{Wang2020,Sirbu2016,Fong2017,Kim2014,Boric2023} and
GameStop \citep{Fong2017,Zhu2016,Zhu2015}. Platforms that support
reviews without selling games include Metacritic \citep{Kasper2019,Fong2017,Straat2017,Cox2015,Kwak2020,Santos2019,Strt2017UsingUC,JeongminSeo2023,SouzaGoncalves2020},
which aggregates reviews, the video game database VideoGameGeek \citep{Koehler2017},
IGN \citep{Fong2017,Zhu2016,Zhu2015,Zagal2009}, GameSpot \citep{Caserman2020,Zhu2016,Zhu2015,Meidl2021,Zagal2009,Bond2009,Zhu2010},
and PC Gamer \citep{Caserman2020}, which are game news portals, and
the serious games portal \citep{Caserman2020}. Reviews in the form
of videos are mentioned less frequently but do occur in dedicated
channels on sites such as Youtube \citep{Ribeiro2019}. Reviews are
also sometimes hosted on blogs and social media sites \citep{Kim2014}
and in the form of comments, for example to game play videos on Youtube
\citep{Silva2022}.

Steam is a popular contemporary option for review analysis because
it is a successful review ecosystem and also because it is possible
to download or scrape large numbers of reviews from this site. Its
size also ensures that even speciality areas are well represented,
such as VR experiences \citep{Gao2022,Faric2019} or reviews in different
languages \citep{Cabellos2022}. Other sites hosting VR experiences
and reviews include Viveport and the Oculus/Meta store \citep{Faric2019,Fagernas2021}.

Academic case studies as a form of review are hosted on publisher
and paper indexing sites, such as the ACM digital library, IEEE Explore,
ISI Web of Science, SCOPUS, Springer Link, Wiley Online, and Google
Scholar \citep{Calderon2015,Giani2016,Petri2017,YanezGomez2016,Zheng2021}.
Review registries and indexing sites are essential in finding relevant
reviews \citep{Bashir2019}. Unlike game distribution ecosystems,
these are often behind paywalls and are not accessible to the general
public or many professions such as classroom teachers. Grey literature
is also a relevant source of valid review information \citep{Calderon2015}
as case studies may also be reported by industry or government sources.

Game reviews were originally published in game focused magazines \citep{Ribeiro2019,Ribbens2012},
such as Computer Gaming World, Computer Games, PC Gamer, Soft World,
Game World \citep{Tsang2009} and MikroBitti \citep{Suominen2011},
and on web portals \citep{Ribbens2012,Tsang2009}. Game reviews for
games in specific categories such as serious and educational games
may be published as collections in books \citep{Coleman2014,Friedrich2019,Barr2017}.
Novel environments include a 3D review environment \citep{Shiratuddin2011}
that is well suited to the review of 3D designs.

\subsubsection{Viability}

The length of time that a review ecosystem has existed (or will continue
to exist) is an indicator of its viability. While the sources consulted
rarely discuss the viability of review ecosystems, they do frequently
describe these systems in terms that refer to a range of indicators
of viability. The major classes of review ecosystem and their viability
indicators are:
\begin{itemize}
\item \textbf{Published~academic~case~studies} \citep{Bas2019,Calderon2015,Giani2016,Petri2017,BRYAND.2006}:
The subtleties of the academic publishing environment are beyond the
scope of this review although it can be regarded as one of the longest-lived
systems with established quality control processes and dedicated,
incentivized contributors. Only particular topics (e.g., usability
evaluation \citep{YanezGomez2016}) vary in popularity over time.
\item \textbf{Game~distribution~platforms}: (e.g., Steam, Apple App Store,
Google Play Store, Oculus/Meta Store and Amazon) have existed without
interruption since their creation and reviews are available for most
of their lifespan \citep{Baowaly2019,Li2021a,Cabellos2022,Fagernas2021,kosmopoulos2020summarizing,Busurkina2020,Wang2021,Tong2021}
making age a predictor of further longevity. While these platforms
focus on distribution they add significant value by also hosting reviews
and integrating these into their business \citep{Guzsvinecz2023}.
This feedback loop ensures all stakeholders benefit from the ecosystem
\citep{Lin2018,Livingston2011,Petrovskaya2022,Kim2014}. It also allows
for quality control by ensuring reviewers have purchased and played
the game \citep{Soetedjo2022}. These platforms use additional viability
indicators: popularity \citep{Fagernas2021,Phillips2021}, number
of products \citep{Urriza2021,Alfaro2021,Yu2021,kosmopoulos2020summarizing,Busurkina2020,Wang2021,Epp2021,Deng2023,Boric2023,alQallawi2021,Chand2022,Yu2023,Philp2024},
market share \citep{Busurkina2020,Wang2021,Philp2024}, turnover \citep{Chand2022,Philp2024},
number of products reviewed \citep{Viggiato2022,Meidl2021}, number
of regular active users \citep{Urriza2021,Yu2021,kosmopoulos2020summarizing,Busurkina2020,Wang2021,Epp2021,Deng2023,Tong2021,Yu2023,Philp2024},
contributions per user \citep{Tong2021}, and number of reviews \citep{Viggiato2022,Guzsvinecz2022,Yu2021,Santos2019,Meidl2021,Boric2023},
with accumulation of information increasing the value of the ecosystem
and range of uses for the reviews \citep{Petrosino2022,Deng2023}.
Viability is adversely affected when applications are removed from
the platform, which may affect their associated reviews \citep{Chand2022}.
\item \textbf{Review~platforms}: (e.g., Metacritic, VideoGameGeek, GameStop,
GameSpot, IGN) achieve longevity by sourcing reviews from both amateurs
and professionals \citep{Kasper2019,Kwak2020}. Additional viability
indicators include: amount of traffic \citep{Zhu2015,Zhu2010}, reputation
\citep{Zhu2015,Cox2015}, amount and relevance of content \citep{Zhu2015,Cox2015},
and diversity of stakeholders \citep{Zhu2015}. Such platforms do
need to adapt to changes in reviewing trends to stay relevant \citep{Koehler2017,Fong2017}
and avoid risks associated with poor quality or fake reviews \citep{Straat2017}.
Perceived commercial conflicts of interest can also affect the value
of the reviews \citep{Zheng2021}. The serious games portal \citep{Caserman2020}
is relatively new and represents a good case study into establishing
a review ecosystem.
\item \textbf{Game journalism}: Media such as magazines and websites with
professional reviewers depend on advertising income and become less
viable when other platforms become more popular \citep{Ribbens2012,Tsang2009}.
The acceptance of games, and game reviews, across society is linked
to reviews spreading from specialist magazines to general purpose
media \citep{Suominen2011}.
\item \textbf{Social media}: Such platforms tend to facilitate comments
rather than formal reviews and so indicate viability through properties
related to the health of the comment stream (e.g., number of commenters
\citep{Silva2022}).
\item \textbf{Professional~review}: Industry processes for quality control
include reviews \citep{Shiratuddin2011}, and are classified as stable
in that these are established professional processes.
\end{itemize}
Custom review processes \citep{Kirschner2014}, or reviews assembled
into books \citep{Coleman2014} are one-off systems for which the
concept of viability as an ecosystem is ill-defined. These are more
viable when using open access allowing them to be easily copied \citep{Barr2017}.

\subsubsection{Challenges}

The previous sections have focused on solutions; strategies and elements
that produce viable review ecosystems. This section focuses on the
challenges; the problems that still need to be resolved in order to
build future review ecosystems.

Consumer sites have demonstrated that it is possible for non-specialist
reviewers to produce reviews that add value to particular communities.
The incentives used (section \ref{subsec:Incentives-1}) ensure that
producing reviews incurs almost zero cost \citep{Ho2012} although
storing and providing access to review information does require costly
expertise \citep{Cho2020}. A key challenge in those environments
\citep{Ho2012} is to verify the motivation of reviewers \citep{Kasper2019,Cox2015}
to ensure that the reviews remain credible and are not manipulated
\citep{Angelis2021,Balietti2016}. Commercial sites may screen reviews
to avoid upsetting advertisers \citep{Tsang2009} while professional
reviewers can experience pressure to score consistently with other
reviews or sales data \citep{Cox2015,Tong2021}. Games are released
in different versions and with updates and reviews need to indicate
clearly which versions are described \citep{Fong2017} so that reviews
are comparable \citep{Cho2020}. The site focus can be a challenge;
for example a game focused site may host VR chat applications but
not facilitate reviews appropriate to these applications \citep{Deng2023}.

Commercial game review sites are satisfied with each review representing
the subjective opinion of a single player \citep{Ribbens2012} unlike
case study reviews where properties such as quality or usability are
aggregated measures generated by sampling significant numbers of players
\citep{Calderon2015}. Having large numbers of people contributing
to a single review is challenging but offers the opportunity to create
robust, reproducible and relevant reviews.

Large numbers of reviews need to be ranked so that only the most relevant
are presented to readers \citep{Baowaly2019}. The correct ranking
of a review may depend on the contents of the review, the preferences
of an individual reader and the evolving needs of the community \citep{Ribbens2012}.
Reviews from the same author are easier to compare, but only professional
reviewers tend to consistently produce multiple reviews \citep{Santos2019}.

Reviews consist of several fields, including a numerical rating for
the product as well as the text of the review justifying this score.
Review ratings provide a single measure that scores the review. This
rating is coarse and introduces the challenge of identifying which
property of the review (if any) is the basis for the rating \citep{Kasper2019}.
Some sites might only provide a binary positive/negative which excludes
even a neutral rating \citep{Faric2019}. Ratings reflect unconscious
bias and lived experience \citep{Kohlburn2022,kosmopoulos2020summarizing}.
Review text tends to consider each game in isolation and may fail
to provide an analytical description of the game play \citep{Zagal2009}
or even a structured comparison listing pros and cons \citep{Boric2023}.
Community review standards adapt to the current state of the games
industry \citep{Ribbens2012} and differ according to cultural values
\citep{Tsang2009}. Further complicating matters, new review ratings
could be interpreted as agreement with previous review ratings \citep{Tong2021}.
Identifying which features of a review are relevant is still an open
problem.

Several review analysis strategies decompose review text into features
which are then analyzed individually. Different contexts may then
utilize only some of these features \citep{Kasper2019,Epp2021}, providing
the challenge of adapting reviews for different purposes such as relevant
issues to each player and specific product information for vendors
\citep{Zhu2015,Yu2021}. Ideally reviews would be structured with
defined sections that support analysis \citep{Viggiato2022}. Formal
review processes provide specific documentation \citep{Shiratuddin2011}
that adheres to defined standards. The diversity of games requires
that new metrics be invented to report on particular properties of
an innovative design \citep{YanezGomez2016,Livingston2010}. An alternative
hypothesis \citep{Koehler2017} is that reviews are holistic and individual
elements should not be considered out of context. This has some support
in that a common language for reviews has yet to be established (see
section \ref{subsec:Template-1}). The challenge is to devise a review
analysis process that does not start with a decomposition stage.

Consistency in reviews, particularly where measures need to be compared
across reviews, requires a systematic process to evaluation \citep{Giani2016}.
The evaluation model needs to be specified in the review, and the
evaluation carried out correctly \citep{Petri2017,WangReview2008}.
Unintuitively, review quality improves when the review is costly to
the reviewer (e.g., requires an investment of effort) \citep{Garcia2017}.
This is a challenge for educational and serious games as there are
logistical challenges to conducting evaluations with students in a
classroom and while being accommodated within an already complex learning
environment. Review quality is preferred over quantity within high-involvement
communities \citep{Zheng2021}. The quality of the game itself is
not sufficient either; reviews of serious games need to report on
how well they meet educational goals as well \citep{Coleman2014}.
Systematic review processes are not currently applied in generating
game reviews \citep{Zheng2021}.

A trend in reviews is to encourage longer detailed reviews since these
are identified as being more useful \citep{Petrosino2022}. Extreme
cases \citep{Kirschner2014} that involve 20 hours of game play, curation
of video recordings and interviews provide significant insights but
with concomitant investment of resources. The challenge is to maximize
value of the review while minimizing the effort required to produce
that review, for example, by focusing only on aspects that would have
the greatest impact \citep{Bashir2019}.

\section{\label{sec:Results}Results}

The detailed analysis leading to these results is presented in section
\ref{sec:Analysis}. This analysis links the sources reviewed to the
categories identified and recommendations made in this section.

\subsection{Purpose}

Review ecosystems exist for a range of purposes. The goal of a new
review ecosystem can be: gaining understanding, sharing insights,
collating information, conducting assessment and promoting products.
The purpose of the review then adapts to this goal by either encouraging
informal reporting by participants, following set processes with groups
of reviewers, or applying consistent and rigorous evaluations across
the entire ecosystem. The information presented in the review may
be intended for later analysis, or designed to be used for a preset
purpose. Opportunities exist to exploit review ecosystems as they
scale, where synergies may result from multiple reviewers contributing
to a common review, reviews covering multiple games, or by aggregating
multiple reviews.

\subsection{\label{subsec:Formofthereview}Form~of~the~review}

The goal is to develop insight into the mechanisms used to create
and manage review ecosystems. This starts by defining the properties
of individual reviews: how they are presented, what information they
contain and what they provide.

\subsubsection{Format}

Reviews are usually written documents that can exist in several forms.
There are few examples of other media formats. A new review ecosystem
would specify the form and format for reviews. Free form text is the
most flexible but complicates analysis. Fixed format documents with
defined sections ensure that the review covers particular topics.
Academic case studies include details of the process used while reviewing.
Opportunities exist to exploit other formats and media; from sets
of numerical ratings to comment streams, images and video, interviews,
streams and recordings of the actual game play, and as reviews embedded
within virtual environments.

\subsubsection{\label{subsec:Template}Template}

A review template specifies particular elements of a review that are
common to all reviews within an ecosystem. Development of a new review
ecosystem is an opportunity to establish a review template that ensures
relevant information is included in each review. Minimum requirements
are product metadata to describe the product being reviewed and the
review content with product description, and pros and cons. Other
template fields are: reviewer details, product specific properties
(for example, games would include: narrative, challenge, gameplay),
quality indicators, and messages to particular groups of readers (e.g.,
developers). Specialist template topics describe educational content,
assessment mechanisms, and the review process. Templates for reviews
of reviews would include review metadata or features extracted from
the reviews.

\subsubsection{\label{subsec:Focus}Focus}

The focus of the review is an opportunity to shape the community that
will form around the review ecosystem. The focus of the review can
be to: inform others, contribute to the community, show status, reflect
on game designs, influence purchases, promote products, provide feedback
to developers, report problems, evaluate educational experiences,
or categorize games.

\subsubsection{Field}

Most reviews analyzed relate to games, with a smaller subset relating
to virtual and augmented reality. Review ecosystems consist of collections
of reviews that provide insight into the domain of the reviews while
having secondary relevance to other overlapping domains. New review
ecosystems should select the class of product to review when building
a review ecosystem. While games are well represented due to the focus
of this $\exper$, sub-fields based on platform (mobile, desktop)
or category of game (multiplayer, serious, simulation, therapy) support
review stakeholders with a particular interest. Areas such as virtual
reality, augmented reality, and mobile applications with their own
sub-fields, can exist independently or within other review ecosystems.
Some studies specialize further to individual dimensions of games
such as play experience, narrative or addiction, or to the intersection
of games and other fields (e.g., health, journalism). Opportunities
exist to identify further fields, or explore the relevance to other
fields of the reviews that exist in established review ecosystems.

\subsection{\label{subsec:Reviewutilization}Review~utilization}

The next step in understanding the motivation that sustains a review
ecosystem is to ask who uses reviews, what they are used for and how
they are used.

\subsubsection{Target~audience}

The audience for a review ecosystem extends beyond just the readers
of the reviews. A new ecosystem can be strategic in targeting particular
audiences. The audience of interactive application reviews can be
one or more of: designers and developers if focused on reasoning about,
and improving, games; customers and publishers when marketing and
selling games; and reviewers and players when discussing games and
the play experience. Game reviews have other audiences: teachers and
students using games to support education, and researchers investigating
reviews and understanding games through reviews. The audience can
extend to include other specialists who want to access games identified
through reviews, e.g., health professionals who can use games for
therapies.

\subsubsection{Usage}

Reviews will find use within their intended focus areas if the review
ecosystem ensures they are useful. This requires strategies that ensure
they are trustworthy and represent the authentic experience of the
author. Useful reviews need to reveal information (e.g., insights
to guide developers) and be available for analysis (e.g., by supplying
many reviews, or by structuring them to simplify information extraction).
Reviews support player communities by describing game play strategies
and offering support for dealing with problems. It must be possible
to identify the relevant and useful parts of a review. Opportunities
exist for review ecosystems to incorporate elements that present and
refine insights based on contributed reviews, and provide this as
a service to stakeholders.

\subsubsection{Education}

Education is a strong theme in a number of the sources identified.
When deciding on the relevance to education of the review ecosystem,
the design choices include: rating and measuring educational value,
identifying and adapting experiences for use in education, and/or
assessing and integrating educational outcomes through creating reviews.
EduXR review ecosystems represent one opportunity to systematically
manage these processes and allow the educational value of different
experiences to be reported and compared.

\subsubsection{\label{subsec:Mechanics}Mechanics}

A review ecosystem supports the creation of reviews that are consistent
with the goals of the system. The processes that achieve this include:
identifying the source of information (one game or many per review;
whether reporting on an experience, assessing a product, or refining
a design; determining the background, professional experience and
sources of bias of the reviewer), enforcing review process rigor (credibility
of the reviewer, experience with gaming and the game, review structures
and standards, levels of objective versus subjective content), providing
efficient review creation tools (ratings, short comments, or long
reviews; tools for analysis as well as reporting), and defining when
the review is performed (aggregation of other reviews before playing,
during play, immediately after play, after significant amounts of
play, long after play based on nostalgia, after watching recordings
of others playing). Opportunities to innovate with these mechanisms
include linking annotated evidence from the game play experience to
relevant sections of the review. This would add value when showing
how to repurpose games for therapies or other purposes.

\subsection{\label{subsec:Ecosystemmanagement}Ecosystem~management}

Review ecosystems create, collate and manage reviews on an ongoing
basis. The interactions between multiple components (the games being
reviewed, the people doing the reviewing, and the environment that
hosts the reviews) is similar to an ecosystem resulting from the interactions
of organisms and their environment. Only a few sources \citep{Koehler2017,Giani2016}
explicitly reason about collections of reviews.

\subsubsection{Structure}

The four categories of ecosystem elements are: the review writers
(communicators), the review users (communicatees), the review hosts
(channels) and the review modifiers. The latter involves the actions
taken after reading the review and other modifications before and
after publication. Of the four ecosystem elements, choices can be
made with respect to: review writers and their purpose (rating, recommending,
repurposing, reasoning or refining the game), the review readers and
their goals (decision making, analyzing experiences to identify the
effects of design decisions, measuring value), review modifiers (translating
reviews so they can be used for additional purposes), and the review
medium (paper, online with restricted access, online and open) with
its associated hosting environment (once-off specialist repository,
curated collections, review site with associated market, specialist
review ecosystem). The review modifiers represent a prominent area
of recent research focused on extracting insight from the readily
available data consisting of large collections of reviews.

\subsubsection{\label{subsec:Incentives}Incentives}

Review ecosystems use incentives to encourage participation. The incentive
mechanisms used in review ecosystems can be significantly enhanced
using innovative practices employed in areas such as gamification
and persuasive technology \citep{Edwards2016,Chin2023,Faraoni2023}.
Traditional incentive mechanisms employed in review ecosystems to
encourage participation (review creation, rating of reviews and reviewers,
community engagement) include: financial incentives, status within
a community, career progression, self-interest including promotion
and marketing products, ecosystem related benefits such as collaboration
and cooperation, emergent ecosystem benefits such as wisdom of crowds,
free stuff where reviews are sources of data for research, scarcity
as an incentive, and opportunities for personal growth). Disincentives
that need to be managed are bias that erodes trust and adverse community
interactions.

\subsubsection{Value}

Stakeholders produce reviews because strategies leading to review
creation have a higher payoff than other strategies. A review ecosystem
maximizes the value of rewards and minimizes the associated costs.
Potential rewards include: intrinsic personal rewards (status, altruism,
improved play experiences), extrinsic personal rewards (money, improvements
in quality of game experiences, ability to identify relevant games),
commercial benefits (improved development processes, higher product
quality, validated targets for evaluating products, increased sales
with less marketing effort), community rewards (support from other
community members, emergent benefits of aggregated reviews) and specialist
rewards (can identify and adapt games for specialist purposes, can
share evaluation tools). Costs that can arise include: overheads associated
with producing reviews (financial cost), overheads associated with
managing the ecosystem (effort of curating reviews, hosting costs),
community management (managing quality and accuracy, setting community
standards), and the effort associated with utilizing reviews (accessing
large sets of reviews, extracting insights from free-form text). An
opportunity for an additional reward is to use reviews to observe
behaviour within gaming communities without adding additional forms
of data collection that might disturb the system.

\subsubsection{Quality}

Review quality measures direct tho reader's attention to relevant
reviews and indicate the value inherent in each review. Measures relate
to individual reviews including: intrinsic measurable review properties
(quality, taxonomy fit, validity), review meta data (reviewer reputation,
sales levels), reviews of reviews (helpfulness) and the ease with
which insights are extracted. Quality is measured for collections
of reviews using aggregation, statistical measures, correlation, ranking
and trend predictions. An opportunity exists to find ways to validate
reviews by including evidence of the claims made.

\subsubsection{Environment}

The review platform can be: an online platform (with product sales
and reviews, dedicated review portal, other review formats such as
video, social media service supporting comments, curated and restricted
platform such as academic publishing, special interest group like
an online gaming community), or printed media (books, magazines).
Beneficial platform services include providing: downloads of reviews
for additional analysis, and effective review ranking and search.
Review environments could make better use of media modalities, with
game play streaming being a form of review, or other platforms (e.g.,
reviews hosted within a game, or virtual environment). Review platforms
that link two different experiences within a single review would facilitate
comparisons and ranking.

\subsubsection{Viability}

The length of time that a review ecosystem has existed (or will continue
to exist) is an indicator of its viability but other measures exist.
When creating a new review ecosystem viability is an indicator of
the longevity of a self-sustaining system as are measures such as
size (e.g., number of reviews, number of users), levels of engagement
(e.g., rate of reviewing, rate of usage), or measurable properties
indirectly related to the reviews such as product sales. Approaches
to achieve viable ecosystems include: combining product sales with
reviews so the feedback loop ensures reviews respond to but also influence
products, paid professional reviewers (e.g., review platforms, academic
research) to ensure review quality and relevance, and integrating
with social media to focus on user engagement. Specialized communities
require continuous injection of resources to remain viable (e.g.,
advertising revenue for journalistic media, professional practices
for quality reviews during development). Opportunities exist for platforms
that facilitate comparison and analysis of reviews during and immediately
after their creation.

\subsubsection{Challenges}

New review ecosystems need to overcome challenges such as: cost (of
creating reviews, of curating them, of validating them), knowing information
about the reviewers (amateur/professional, motivation, experience,
reputation), review process and structure (one contributor per review
or many, formal or ad hoc reviewing processes, defined topics or free
text, numerical rankings and/or qualitative opinions, one game per
review or many, long detailed reviews or streams of short comments),
and utilization (rankings for reviews, sharing, analysis processes,
reuse and repurposing). Some of these combinations represent new opportunities
such as challenging the notion of a review as a single document produced
by a single author representing a single instant of the experience.
For example, there are proposals for $\scoping$s which receive continuous
updates from multiple contributors rather than generating duplicates
with minor refinements \citep{Arksey2005,Shojania2007,Ioannis2016}.

\section{Discussion}

The \textbf{first goal} of this review is to identify and describe
trends related to current practices in preparing, presenting and maintaining
a set of $\exper$s within an $\exper$ ecosystem. The analysis (section
\ref{sec:Analysis}) and results (section \ref{sec:Results}) extract
and summarize the strategies used within known ecosystems within the
broad headings of form, utility and ecosystem. The \emph{form} of
an $\exper$ determines what information is included and how it is
presented (section \ref{subsec:Formofthereview}). This then leads
into considering the \emph{utility} of an $\exper$ and the value
that it needs to provide (section \ref{subsec:Reviewutilization}).
The community that develops around an \emph{$\exper$ ecosystem} provides
the value for all participants, and completes the feedback loop of
producing and consuming $\exper$s that ensures the ongoing stability
of the system (section \ref{subsec:Ecosystemmanagement}).

The \textbf{second goal} is to present practices that can be used
to establish and enhance the $\exper$ ecosystem for eduXR experiences.
This outcome is summarized in Figure \ref{fig:Guidelines-for-the}.

The following sections discusses trends identified through the relationships
between each topic and present the opportunities to establish new
ecosystems that have been identified through this synthesis.

\subsection{Synthesis}

\subsubsection{Form}

Figure~\ref{tab:Template-1} identifies the categories of information
present in $\exper$s. Review ecosystems for eduXR would extend these
to identify and include information that captures the intersection
of education and XR. Additional information would expand on the quality
of the experience, the social community, the development and commercial
opportunities, and diverse specialist needs such as education and
therapy. The amount of time spent playing is a valuable indicator
of review quality. Alternative measures are required for eduXR experiences
that are of fixed duration, or have a linear narrative that discourages
replay, to accommodate the new cohorts of students experiencing them
for the first time. An XR experience can include game play but can
also be very different in how it makes use of the reality spanning
medium and the value that it provides to participants.

Despite the growing popularity of game streaming services as a form
of video review the format of reviews reported are written documents
and mostly unstructured content as free form text. Attempts to structure
reviews are common but no mechanism has yet been identified to ensure
a single standard is adhered to or that such would meet needs beyond
those of niche groups. Natural language processing can extract topics
from review text and can mine reviews for insights beyond those expected
from a fixed review template. Reviews recorded directly in XR would
capture the experience directly but would add challenges related to
utilizing the review. Regardless of the review format, they need to
be indexable to support searching and comparisons of reviews. Review
presentation can be separated from the form used for review creation
with opportunities to restructure and reformat, for example, to present
a customized overview summary with highlights using multiple media.

\subsubsection{Utility}

Reviews attract a diverse audience but the nature of the reviewer
themselves is often overlooked. Information about the reviewer provides
additional insights, authenticity and utilizes reputations. Review
systems that employ social network mechanics share reviewer details
and profiles. Academic $\exper$s reveal personal author information
but also details of the number of users who tested the experience
and their demographics. Reviews can be prepared directly by those
who use the experience (e.g., expert reviewers, players who write
their own reviews) or distilled from reports of others (e.g., user
questionnaires). An assumption underlying most review systems that
a game player role intersects with the review author role. Recognition
of roles leads to opportunities to explicitly include information
describing the person behind each part of the review and envisage
additional roles. For example, an $\exper$ need not only report on
the player's experience but could separate the stages of play (or
generating evidence or provocations) from the analysis which could
be performed by specialists across several fields as pioneered by
\citep{Kirschner2014}.

Reviews are distinctive in that they are trusted. While the mechanics
of trust are complex, trust is typically achieved through the alignment
between the interests of the reviewer and the consumer of the review.
Potential for bias is also reduced when there are multiple roles contributing
to a single review.

Analysis and data mining identifies patterns present across an entire
repository of reviews. Review ecosystems tend to assume that review
is used in the way that the author intends it to be. In the context
of an eduXR ecosystem we anticipate that aspects of a single experience
may be adapted for use across several lessons with different learning
outcomes. Different portions of a review may hint at these different
opportunities. Review ecosystem utility increases when further insights
can be extracted from existing reviews, or if review structures can
adapt to increase the value that can be extracted.

\subsubsection{Ecosystem}

Review ecosystems are most prominent in the platforms that host and
sell games and use reviews to guide player purchasing decisions. The
secondary ecosystem around educational games uses reviews to select,
but also validate, the educational value in a game. Academic research,
such as case studies, also generates product reviews with added rigour
applied to the method and with a focus specific to a field of research.
Internal review ecosystems within organizations mirror the structure
of other ecosystems with explicit goals for the review process. Creating
a review ecosystem starts with defining goals. The ecosystem itself
is then structured around the components that write reviews, use reviews,
host or communicate reviews, and that modify and extract insights
from reviews.

The incentive mechanisms used in review ecosystems are rudimentary
compared to what could be achieved. This may be a constraint of the
$\scoping$ methodology since sophisticated incentive and behavioural
manipulation strategies employed for commercial benefit are often
not discussed publicly. Games already employ a wide range of incentive
mechanisms to keep the player engaged that can be extended to gamify
behaviour consistent with a stable and valuable review ecosystem.
Academic research and publishing is another ecosystem with its own
set of incentives (rewards and costs). Regardless of the merits of
this system, its stability suggests analogs of the incentives used
could be adapted to an eduXR review ecosystem. However, the trust
associated with a review ecosystem lies in the honesty of the incentive
mechanisms that encourage voluntary contributions, personal growth
and community values. Without these, removal of biased reviews requires
investment of considerable resources.

Feedback loops are also a form of control system for regulating an
ecosystem. Ecosystem mechanics balance the costs of contributing with
incentives to ensure a stable ecosystem without opportunities for
subversion. Incentives can be explicit rewards or implicit incentives
such as social status. A review of reviews is a feedback cycle where
review writers are able to identify the value of their work and improve
their reports even in the absence of an explicit quality measure.
Quality metrics measure both properties of the XR experience but are
also used as a filter to identify relevant $\exper$s. The numerous
quality related properties of interactive and educational experiences
listed in Figure~\ref{tab:Template-1} will be extended as additional
eduXR experiences are developed. Quality measures are often proxy-measures
of other quantities (e.g., agreement, reputation) and are too coarse
to identify specific actions to take in response. Manually assigned
scores are subjective, prone to variation between individuals, and
can be manipulated by social network manipulation (e.g., memes, review
bombing). Automated text analysis trained on manually assigned scores
may inherit their bias. Review quality needs to be managed where there
are conflicts of interest or sources of bias. A known model that predicts
quality can be attacked by procedurally generating reviews with these
properties or other adversarial strategies. An $\exper$ is itself
a quality measure that trades detail against its flexibility in comparing
different experiences. A universal measure of quality may be an aspirational
goal but quality measures customized to each stakeholder would provide
equivalent value. These would also be used to provide feedback to
authors when creating reviews.

\subsection{\label{subsec:Case-study}Case study: eduXR review ecosystem}

To illustrate how this review achieves the second goal related to
establishing new review ecosystems, the concepts distilled in this
paper are applied to the challenge of developing a review ecosystem
suited to the specific needs of teachers as described in section \ref{subsec:Rationale}.
The alternative strategies summarized in Figure \ref{fig:Guidelines-for-the}
are matched to each requirement to provide a high-level design proposal:

\emph{Identify relevant experiences}: The \emph{review ecosystem}
must focus on consistent evaluations whose purpose is know apriori.
The \emph{target audience} is teachers and students with opportunities
to provide feedback to developers who can enhance experiences to teach
particular topics. The \emph{mechanisms} focus on rigor by requiring
objective information, standardized review processes and reporting
on the reviewer's credibility. The \emph{incentive} for review creation
is that the review ecosystem return value to authors by providing
access to reviews created by others. Reputation established and tracked
within the review ecosystem also provides \emph{value} for the contributors.

\emph{Efficiently accessing information}: A written text \emph{format}
is efficient to access and search combined with images and video for
a quick preview showing the application in action. The review document
uses a \emph{template} with fields providing information on how the
experience can be applied to particular curriculum topics. Enforced
templates and formatting requirements are facilitated by hosting the
review ecosystem in an online \emph{platform} that can verify these
constraints as the review is created.

\emph{Extract details relevant to teachers}: \emph{Education}al purpose
involves assessing experiences for educational value and providing
suggestions on how to adapt experiences to resource constrained classrooms.
\emph{Quality} measures focus on educational insights and ease of
applying information from the review. Review ratings should explicitly
measure usefulness of the details provided, which are also relevant
when searching for information.

\emph{Focused on educational elements}: The \emph{focus} of the review
is to support the community of educators by collecting and presenting
information about the educational properties of the experiences. While
the \emph{field} addressed by the review is education, this must include
sub-fields corresponding to different topics that are taught.

\emph{Validated information}: The \emph{usage} of the reviews requires
relevant, up to date, and trustworthy information. The ecosystem supports
this by presenting details of the evaluation process and reputation
of the reviewer. Reputation mechanisms reduce \emph{costs} associated
with screening and validating reviews. The ecosystem \emph{structure}
has both review writers and users drawn from the teacher population
to ensure relevant information is communicated.

This study identifies that the requirements have left out the \emph{challenges}
associated with the review ecosystem. \emph{Viability} can only be
achieved if the \emph{costs} are managed and if the content remains
up to date. This may require including product marketing where there
are benefits of having product vendors both promote and facilitate
reviews of their products, provided the trustworthiness of the review
ecosystem can be maintained. Bootstrapping the system is a challenge
where the new platform could be seeded with insights automatically
extracted from reviews in existing ecosystems.

These strategies are adapted from those used in existing review ecosystems
(see section \ref{sec:Analysis} for a summary of these). There are
opportunities to experiment with innovative approaches specific to
eduXR. Having multiple reviewers able to contribute to a single review
would ensure that the review represents a greater consensus, incorporates
a wider range of discipline expertise, reduces the number of reviews
returned by a search, and reduces the effort required by individual
reviewers.

\subsection{Limitations}

The concept of a review ecosystem for interactive experiences is not
yet well established in academic literature. This is further complicated
by having the word ``review'' being very common in academic databases.
We ensure that all relevant sources of information are covered using
the two phase search strategy that starts with a broad search combined
with manual screening. The second phase then uses focused search strings
that returns a dense set of relevant results that are checked for
saturation. There are a range of existing and emerging game and XR
review sites that have not been the subject of academic studies. These
can offer further insights into emerging review ecosystems but will
not be identifiable by a $\scoping$ search methodology. Review ecosystems
focused on supporting commercial goals through product reviews is
also outside our scope.

Research into measures that indicate the quality of an experience
\emph{during participation} have been excluded. This is a potential
useful area to explore with respect to XR applications since these
increasingly have the ability to measure engagement during the experience
using body tracking and facilities such as facial and eye tracking
being included in more recent devices. Analysis and aggregation of
such measures would be particularly relevant to selecting effective
eduXR experiences.

This paper investigates the opportunity for a technology transfer
from review ecosystems for games to other areas such as eduXR. Established
solutions exist for games but the concepts involved may not translate
directly to eduXR. The findings only provide a basis for the deliberate
design of a stable and useful eduXR review ecosystem where it is possible
to adapt the insights gained from game review ecosystems.

\subsection{\label{subsec:Opportunities}Opportunities}

\begin{figure*}
\begin{centering}
\includegraphics[width=0.7\paperwidth]{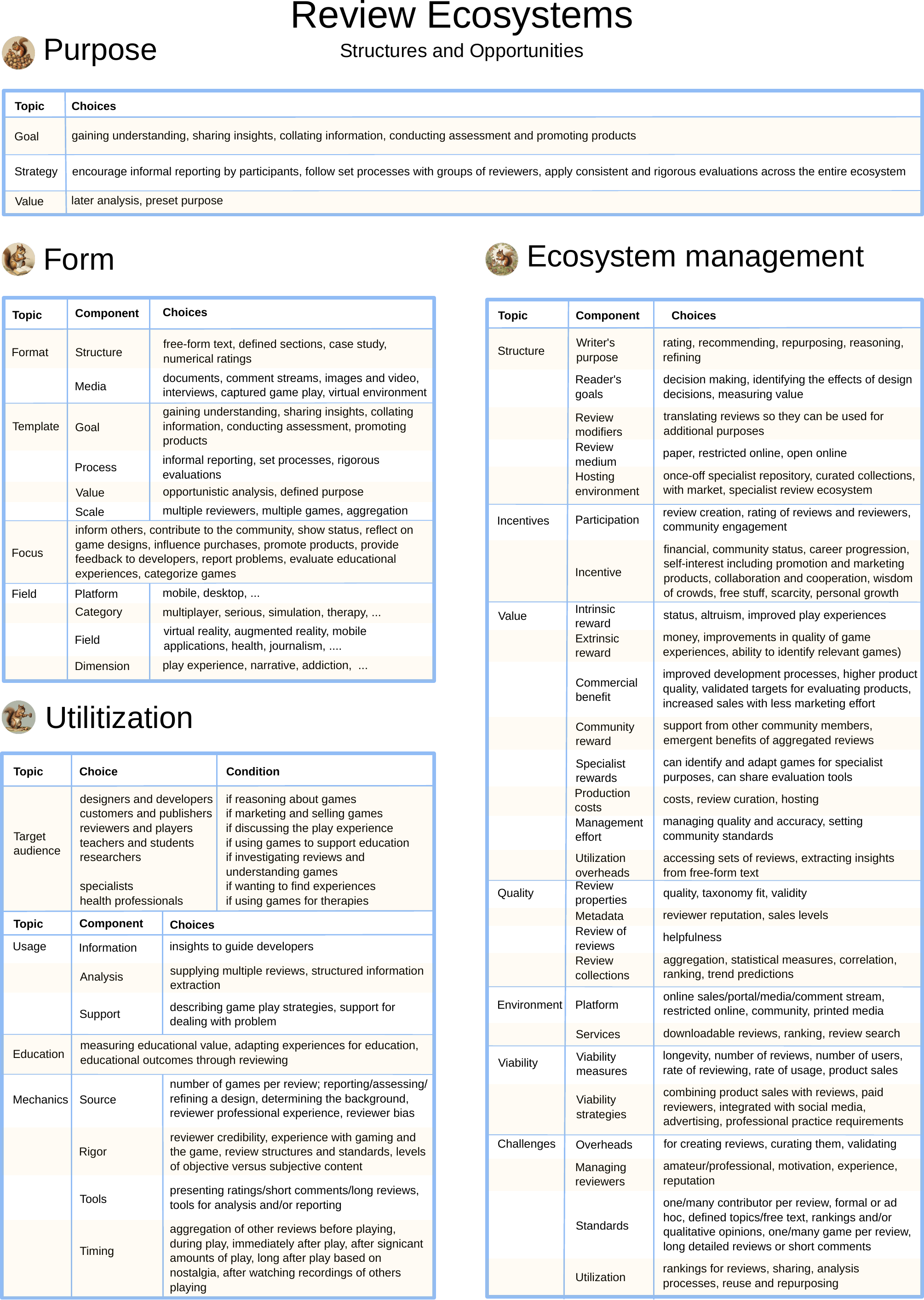}
\par\end{centering}
\caption{\label{fig:Guidelines-for-the}Guidelines for the construction of
review ecosystems.}
\end{figure*}

The analysis (see section \ref{sec:Analysis}) leads to the synthesis
of a set of options that can be considered when building new review
ecosystems. These are summarized in Figure \ref{fig:Guidelines-for-the}.
 However, the choices that do not exist in any existing system
provide opportunities for further innovation with respect to review
ecosystems.

The stage at which an $\exper$ is created (before the experience
based on other reviews, during development of the experience, during
or after the experience, immediately after, or on reviewing recordings)
shapes the review ecosystem. An $\exper$ is not a standalone entity.
Each review is influenced and informed by different factors including
past experiences, past reviews read, how often you've played the game,
how experienced you are as a gamer, your profession, and your motivation
for creating the review. It is only when you know this contextualizing
information that you can interpret the review in a meaningful way.
Since everybody sees the world differently, the process for creating
an $\exper$, including whether certain contextualizing information
be included, can emphasize consistency in communicating concepts,
uncover aspects that may be assumed to be obvious, or present an interpretation
of the experience as understood by that participant. A review ecosystem
is a mechanism to collate diverse opinions to yield collective insight
and these options can be selectively varied to account for a particular
perspective.

The $\exper$ ecosystems are diverse despite the focused keywords
used to identify the sources for this $\scoping$. This variety applies
to the form of a review, the ways in which they are used and the motivation
to create and consume them. Even so, an $\exper$ is still part of
an even larger system when considered within a socio-cultural-historical
perspective \citep{Rieber1997}. Reviews have the unique property
of providing a proxy for an experience. The influences that affect
the creation of an $\exper$ extend beyond the review ecosystem to
include the broader background of the reviewer, their society and
culture, and their context at the time of writing. Existing review
ecosystems do not capture this background even where it would be relevant
in selecting and interpreting $\exper$s. For example, knowing that
a reviewer is an experienced educator, with a long history in trialing
or using digital technologies in the classroom, would be relevant
in assessing the review of an XR experience (with respect to teaching
a particular topic) for readers in similar occupations. The associated
opportunity is: capture and present context relating to the reviewer
and their influences to support interpretation of what is in the review
(and what might be excluded).

The $\scoping$s cited \citep{Petri2017,YanezGomez2016,Calderon2015,Zheng2021}
focus directly on the analysis of existing games and XR experiences
but also set the precedent for analysing existing academic case studies
to perform a $\scoping$ of $\exper$s. Analysis of game reviews using,
for example, data mining works across a broad set of reviews with
the goal of measuring properties of $\exper$s. An opportunity exists:
use $\scoping$ methodologies applied to $\exper$s to identify trends
across classes of game (e.g., in the same genre, or representing evolution
over time of game designs or mechanics). This introduces challenges
around the reliability of $\exper$s as they represent a form of gray
literature, but offers value in providing rigour within review ecosystems
and applying evidence based direction to the design of eduXR experiences.
Computing educators are well positioned to anticipate and exploit
the value that arises from access to a well-structured data set. Reviews
are a tool that is used for many purposes, beyond just describing
and recommending, and can support search and selection, compare properties
and features, describe how to adapt and repurpose experiences for
other purposes, provide feedback and quality control, give insight
into an experience and provide personal growth and connections.

The complexity of a general purpose review ecosystem is a challenge
when constructing a new one. Specialist areas, such as eduXR, would
be best advised to focus on a single core focus across each of the
dimensions (as in section \ref{subsec:Case-study}). The classification
of review ecosystems provides further value by enabling categorization
according to the classification scheme in Figure \ref{fig:Guidelines-for-the}.
This provides a language to describe and reason about existing systems
in order to identify suitable systems for particular purposes. It
allows opportunities to identify systems that provide part or all
of the functionality of a review ecosystem. For example, the academic
publishing environment has been shown to be a review ecosystem in
this paper.

The focus in this paper is on being able to identify and employ an
ecosystem of $\exper$s specifically focused on eduXR experiences.
Mature environments exist for gaming reviews. These can incorporate
eduXR experiences although bespoke communities are starting to emerge
in this category. The significant opportunity resulting from this
study is to: create an eduXR $\exper$ ecosystem that captures and
presents relevant information, provides the insights required to select
and deploy eduXR experiences, and ensures that a strong social community
is built with ethical incentive mechanisms that benefit all stakeholders.

\subsection{XR Review Ecosystems}

XR reviews are currently treated as part of existing game review ecosystems.
However, XR has some fundamental differences, as illustrated in Figure
\ref{fig:The-overlap-between}. The design strategies for XR review
ecosystems are based on a range of existing review systems, including
product reviews, peer review and gaming review communities. The resulting
review guidelines (Figure \ref{fig:Guidelines-for-the}) also reflect
the opportunities that have been identified for XR. The form of the
review extends beyond text and video to capture immersive representations.
Review structure includes concepts from game play but extends these
to other templates that include fields relevant to areas like eduXR
(e.g., curriculum and assessment). The utilization of XR reviews deals
with more than content consumption and entertainment. Reviews for
XR are particularly relevant to areas of training and tourism. Specialist
XR systems connect stakeholders, such as educators to developers and
publishers, as XR applications can be relevant to multiple markets.
Social XR applications lend themselves to collaborative review review
structures, with immersion further developing trust.

This paper does not prescribe a single XR review ecosystem but presents
a systematic approach to creating bespoke environments intended for
specific purposes. This work enables further focused research into
innovative strategies such as collaborative reviews, within-experience
reviews, and customized feedback loops between stakeholders.

\subsection{Contributions}

This paper collates concepts from research into reviewing strategies
for use as components of a complex emergent system. This exposes issues
such as stakeholder roles, varied incentive mechanisms, choice of
evaluation metric and structures for feedback loops. The collected
strategies and options can now be used for purposefully creating review
ecosystems. This framework did not exist previously, resulting in
many electronic platforms still replicating the text heavy, archive
style review environments derived from paper-based publishing. All
options are linked to their underlying research source (see Appendix
\ref{sec:Analysis}). Opportunities afforded by considering reviews
at a system level are identified and discussed throughout sections
\ref{sec:Results} and \ref{subsec:Opportunities}. These result from
the analysis which considers not only each component, but also the
potential interactions between the components.

This work focuses on review structures suited to XR and eduXR, supported
by the case study in section 4.2. The use of reviews to efficiently
find and adapt XR experiences to support teaching is a particular
focus. The scope of a review ecosystem extends past the sales focus
of commercial ratings systems to include many aspects of the community
focused review systems developed for games. XR can be a medium for
communicating reviews, can provide ways to collate, present and evaluate
information relevant to XR, and opens opportunities to pioneer the
inclusion of a review ecosystem as part of the consumption of XR experiences.

\section{Conclusions}

This paper presents a scoping review that catalogs existing practices
in preparing, presenting and maintaining a set of $\exper$s in an
$\exper$ ecosystem. Given the lack of prior research into eduXR review
ecosystems, this process systematically identified relevant literature
from the adjacent field of game reviews. Each source is analyzed with
respect to a range of criteria within the categories of form, utilization
and ecosystem management. A detailed analysis is provided in section
\ref{sec:Analysis} with the resulting components and design choices
listed in section \ref{sec:Results}. This review achieves two goals:
\begin{enumerate}
\item Identify and describe trends related to $\exper$ ecosystems: The
review draws insight primarily from established game review ecosystems
and also considers novel variations of these, as well as review systems
within education and academia. Popular strategies that integrate review
systems into other platforms benefit from the resulting social structures,
use structured text based reviews suited to data mining, and avoid
conflicts of interest that damage the vital ingredient of trust. Opportunities
for innovation include devising measures of review quality that provide
appropriate incentives, improve rigour and value through multiple
reviewers or comparative reviews, integrate the review with the interactive
experience to provide supporting evidence, and connect even more pairs
of stakeholder roles with mutually beneficial relationship structures.
\item Present practices for establishing $\exper$ ecosystems: Section \ref{subsec:Case-study}
illustrates the process for designing a custom ecosystem based on
particular requirements. Figure \ref{fig:Guidelines-for-the} provides
the categories to be considered and the choices to be made. Examples
of the practice associated with each choice are provided through the
citations linked the corresponding tables and discussions in sections
\ref{sec:Analysis} and \ref{sec:Results}.
\end{enumerate}
The categorization of eduXR review ecosystems in this paper provides
a tool for reasoning about such systems. Existing approaches can be
classified according the criteria and options shown in Figure \ref{fig:Guidelines-for-the}.
The merits of different strategies can be compared within each category
and the alternative options presented can be considered as strategies
to refine and improve existing systems, and to support the design
and development of new review eduXR ecosystems.

This $\scoping$ shows that there is only a small amount of existing
research into review ecosystems \citep{Koehler2017,Urriza2021,Bashir2019,Zheng2021},
and that these ideas have not been significantly developed. While
automated analysis of existing review ecosystems is common, this paper
represents the first steps in presenting strategies for deliberately
constructing review ecosystems to facilitate their collective use.

\section*{Declarations}

\subsubsection*{Funding and Competing Interests}

This research received no specific grant from any funding agency in
the public, commercial, or not-for-profit sectors. The authors have
no financial or proprietary interests in any material discussed in
this article.

\subsubsection*{Data availability}

All data used is sourced from the references cited. Copies of the
analysis spreadsheet are available from the authors upon reasonable
request.

\bibliographystyle{IEEEtran}
\bibliography{/mountpoints/bigA/svn/papers/BibRef/RefBase}

\end{document}